\documentclass[a4paper,11pt]{article}
\usepackage[english]{babel}
\usepackage[latin3]{inputenc}
\usepackage[T1]{fontenc}
\usepackage{textcomp}
\usepackage{epsfig}
\usepackage{latexsym}
\usepackage{graphicx}
\usepackage{amsfonts,amssymb,bm}
\usepackage{color}
\newtheorem{proposition}{Proposition}
\newtheorem{lemma}{Lemma}
\newtheorem{theorem}{Theorem}

\newcommand{\beq}{\begin{equation}}

\newcounter{llista}

\begin{document}
\title{Flat deformation of a spacetime admitting two Killing fields.}
\author{Josep Llosa$^1$, Jaume Carot$^2$\\
\small $^1$ Departament de Física Fonamental, Universitat de
Barcelona, Spain\\ \small $^2$ Departament de Física, Universitat de
les Illes Balears, Spain}

\maketitle

\begin{abstract}
\noindent It is shown that, given an analytic Lorentzian metric on a 4-manifold, $g_{ab}$, which admits two Killing vector fields, it exists a local deformation law $\eta_{ab} = a \,g_{ab} + b\,H_{ab}$, where $H_{ab}$ is a 2-dimensional projector, such that $\eta_{ab}$ is flat and admits the same Killing vectors. We also characterize the particular case when the projector $H_{ab}$ coincides with the quotient metric. We apply some of our results to general stationary axisymmetric spacetimes\\[1ex]

Mathematics Subject Clasification: 83C20, 83C15, 53B30
\end{abstract}

\section{Introduction \label{S1}}
It has been recently proved \cite{LlosaSoler05},\cite{CarotLlosa08} that given a semi-Riemannian analytic metric $g_{ab}$ on a manifold $\mathcal{M}$, locally there exist two scalars $a$ and $b$ and a 2-dimensional projector $H_{ab}$ ---i.e. $H_{ab}g^{bc} H_{cd}=H_{ad}$  and $(H_{ab}g^{ab})=2$--- such that the deformed metric
\begin{equation}  \label{e0}
\eta_{ab} := a g_{ab} + b H_{ab}
\end{equation}
is flat. We call this formula the {\em deformation law} associated with $(a,\,b,\,H_{ab})$.

The 2-dimensional projector $H_{ab}$ defines an almost-product structure \cite{Kobayashi},\cite{Ferrando04} $g_{ab}=H_{ab}+K_{ab}$ and the deformation law (\ref{e0}) differently scales the plane $H_{ab}$, by a factor $\varphi=a+b$, and the plane $K_{ab}$, by a factor $a$.

We also proved in ref. \cite{CarotLlosa08} that in case that $g_{ab}$ admits a Killing field $X^a$, then a deformation law can be found such that $\eta_{ab}$ also admits $X^a$ as a Killing field.

Assume now that $g_{ab}$ admits a wider Lie algebra $\mathcal{G}$ of Killing fields, i.e. dim$\mathcal{G}> 1$. Is it possible to find a deformation law such that $\eta_{ab}$ admits any $X\in \mathcal{G}$ as a Killing field? Notice that, as $\eta_{ab}$ is flat, the Lie algebra of its isometries is maximal, i. e. the Poincaré algebra $\mathcal{P}$. Therefore, in order that the answer to the question above is affirmative, it is necessary that $\mathcal{G} \subset \mathcal{P}$. We thus advance the following conjecture:
\begin{quote}
If $g_{ab}$ admits a Killing algebra $\mathcal{G}$ and $\mathcal{G} \subset\mathcal{P}$, then there exist deformation laws such as (\ref{e0}) such that $\eta_{ab}$ is flat and any $X \in \mathcal{G}$ is a Killing field of $\eta_{ab}$
\end{quote}

The problem we shall tackle in the present paper is a little bit simpler: we shall consider an analytic\footnote{As the Cauchy-Kovalewski theorem is invoked at some point in the proof, the validity of the results presented here is restricted to the analytic cathegory} metric $g_{ab}$ admitting two commuting Killing fields and we shall see that a deformation law (\ref{e0}) exists such that $\eta_{ab}$ is flat and admits the same Killing fields. We shall confine ourselves to the case in which the metric induced on the plane spanned by the Killing vectors is non-degenerate and hyperbolic, that is: timelike orbits (whereas the elliptic case can be dealt in a similar way, the degenerate case is quite different). The paper is structured as follows: in section \ref{S3} we present the formalism and prove some intermediate results\footnote{With a different notation, this formalism was developed in \cite{Geroch72} and we present it here in a way suited to our purposes} which we sall apply in section \ref{S5} to prove the stated result. The formalism allows a reformulation of the proof in the quotient 2-manifold, so that a dimensional reduccion occurs. In section \ref{S6} we study the specially simple case when the almost-product structure implicit in the deformation law coincides with the almost-product structure induced by the Killing fields and we apply the above results to the case of stationary axisymmetric spacetimes.

\section{Spacetimes admiting two commuting Killing vectors \label{S3}}
Let ${\cal M}$ be a spacetime with a metric $\eta_{ab}$ admitting two commuting Killing vectors $X_A^a$,
$$ \mathcal{L}_{X_A} \eta_{ab}=0 \,, \qquad A=1,2. $$
Note that, at this point, $\eta_{ab}$ does not designate necessarily a flat metric.
Through any point $x \in {\cal M}$ there is an integral submanifold ${\cal V}_x$, i.e. $T_x{\cal V}_x = {\rm span}\{X_A^a,\, A=1,2\}$, which we call {\em the orbit} trough $x$. Let
$\{e_a,\,a=1 \ldots 4\}$ be a base in $T_x{\cal M}$ and $\{\omega^b,\,b=1 \ldots 4\}$ its dual base.
We denote by $\lambda_{AB}$ the metric products:
\begin{equation} \label{e1}
\lambda_{AB} := X_A^a \xi_{B\,a}\,, \qquad {\rm and}\qquad \lambda^{AB} \lambda_{BD} = \delta^A_D \,, \qquad \mbox{ where }
\qquad \xi_{Aa}:=\eta_{ab} X_A^b
\end{equation}
and define
$$ \xi^{A\,a} := \lambda^{AB} X_B^a   \qquad {\rm and} \qquad X_B^a \xi^A_a = \delta^A_B $$
that is, $\xi^A_a,\, A=1,2$ is the dual base for $X_A^a,\, A=1,2$, on $T_x{\cal V}_x$.

The metric induced by $\eta_{ab}$ on the orbit ${\cal V}_x$ is $\lambda_{AB} \xi^A_a\xi^B_b = \xi_{A\,a}\xi^A_b$ which, as already mentioned, will be assumed hyperbolic, that is,
\begin{equation} \label{e2}
\det(\lambda_{AB})=:-\frac{\tau^2}{ 2} < 0
\end{equation}
and that, in a terminology borrowed from principal bundles theory \cite{Choquet}, we shall call {\em the vertical metric}. Note that, from its definition and the fact that the Killing vectors commute, it is immediate to see that $\lambda_{AB}$ is preserved by $X_A$, i. e. $\mathcal{L}_{X_A} \lambda_{AB} =0$.


Let us assume that the set of all orbits of $X_A^a$, $A=1,2$, is a 2-manifold, i. e. the quotient manifold ${\cal S}$.
The tensor
\begin{equation} \label{e3}
 h^a_b :=\delta^a_b -X_A^a \xi^A_b
\end{equation}
projects then vectors in $T{\cal M}$ onto vectors that are orthogonal to the orbits. Again, in analogy with the principal bundles terminology, vectors that are orthogonal to the orbits will be called {\em horizontal}.

There is a one-to-one correspondence \cite{Geroch71} between tensor fields $T^{\prime\,a\ldots}_{b\ldots}$ on ${\cal S}$ and horizontal tensor fields  on ${\cal M}$, i. e. those $T^{a\ldots}_{b\ldots}$ fulfilling
\begin{equation} \label{e4}
 X_B^b T^{a\ldots}_{b\ldots} = 0 \,, \qquad \xi^A_a T^{a\ldots}_{b\ldots}=0 \qquad {\rm and} \qquad {\mathcal{L}}_{X_A} T^{a\ldots}_{b\ldots}=0 \,, \quad A,B=1,2
\end{equation}
that is, tensor fields that are horizontal and  Lie-constant along $X_A^a$. Following Geroch \cite{Geroch71} \guillemotleft  While it is useful conceptually to have the two-dimensional manifold $\mathcal{S}$, it plays no further logical role in the formalism. We shall hereafter drop the primes: we shall continue to speak of tensor fields being \emph{on}
$\mathcal{S}$, merely as  a shorthand way of saying that the field (formally, on $\mathcal{M}$) satisfies (\ref{e4})\guillemotright

From $\mathcal{L}_{X_A}\eta_{ab}=0$ and the fact that the Killing vectors commute, it follows that the horizontal metric
\begin{equation} \label{e5}
 h_{ab} :=\eta_{ab} - \xi_{A\,a} \xi^A_b \qquad \mathrm{i.\;e.}\qquad h_{ab} = \eta_{ac}h^c_b  
\end{equation}
is preserved by $X_A$, $A=1,2$. By the above mentioned correspondence, it induces a metric on ${\cal S}$ and, as the vertical metric is hyperbolic, $h_{ab}$ is elliptic. We shall denote the \emph{inverse horizontal metric} as $\displaystyle{h^{ab}}$, one then has $\displaystyle{h^{ab} :=\eta^{ab} -\xi^{A\,a} X_A^b}$  and  $h^{ab} h_{bc} = h^a_c$.

\subsection{The Killing equation \label{SS3.1}}
From ${\mathcal{L}}_{X_A}\eta_{ab}=0$ it follows that $\nabla_a\xi_{A\,b}$ is skewsymmetric, or
\begin{equation}  \label{e6}
\nabla_a\xi_{A\,b} = \frac12\,\left(d\xi_A\right)_{ab}
\end{equation}

As the Killing vectors commute, it follows easily that:
\begin{equation} \label{e7}
X_A^a \lambda_{BC|a} = 0 \qquad  {\rm and} \qquad X_A^a \tau_a = 0 \,,
\end{equation}
where $d\tau =\tau_a\omega^a$ and a stroke $|$ denotes differentiation. Hence, $\lambda_{BC}$ and $\tau$ are functions on ${\cal S}$ and  $\lambda_{BC|a}, \, \tau_b$ are 1-forms on ${\cal S}$.

A further consequence of (\ref{e7}) and the commutativity of the Killing vectors is that $\mathcal{L}_{X_B} \xi^A_a = 0$, whence it follows that
\begin{equation} \label{e7a}
i(X_B)d\xi^A = 0\qquad {\rm and} \qquad \mathcal{L}_{X_B} d\xi^A_a = 0 \,, \qquad {\rm therefore,} \qquad d\xi^A \in \Lambda^2{\cal S} \,.
\end{equation}
As ${\cal S}$ has only 2 dimensions, $d\xi^A = \theta^A\epsilon$, where $\epsilon$ is the volume tensor (see Appendix A) and we call the scalar $\theta^A$, the {\em twist} of $X_A$. Then, including (\ref{e6}), we have that
\begin{equation}  \label{e8}
\nabla_c \xi_{A\,d} = \lambda_{AB\,|[c}\xi^B_{d]} + \frac12\,\theta_A \epsilon_{cd}  \qquad  {\rm and}  \qquad
\nabla_c \xi^{A}_d =  -\lambda^{AD}\lambda_{\,DB|(d} \xi^B_{c)} + \frac12\,\theta^A \epsilon_{cd}
\end{equation}
with $\theta_A := \lambda_{AB}\theta^B$.

\subsection{The Levi-Civita connection on the quotient manifold ${\cal S}$ and the Riemann tensor \label{SS3.2}}
Given a horizontal tensor $T^{a\ldots}_{b\ldots}$ we define
$$D_c T^{a\ldots}_{b\ldots} := h_c^d \,h^a_e\,h^f_b\, \nabla_d  T^{e\ldots}_{f\ldots}\,.$$
From (\ref{a5}), it follows that for a horizontal vector $w^b$:
\begin{equation} \label{e9}
D_a w^b = \nabla_a w^b + w^d\,\left[-\frac12\, X_A^A \lambda_{AB|d}\xi^{B\,b} + \frac12\, \theta_A \,\left(\xi^A_{a}\epsilon^b_{\;\;d} + \xi^{A\,b}\epsilon_{ad} \right)\right]
\end{equation}

It can be easily proved that $D$ is a symmetric linear connection on the quotient manifold. Moreover, since $D_a h_{bc}=0$, it is the Levi-Civita connection for $h_{bc}$ in ${\cal S}$.

The Riemann tensor ${\cal R}_{cdab}$ for the connection $D$ can be derived from the Ricci identities, $D_a D_b w^c - D_b D_a w^c= w^d {\cal R}^c_{\;dab}$, and one thus  obtains
\begin{equation} \label{e10}
{\cal R}_{dcab} = R^\perp_{dcab} - \frac12\,\theta_A \theta^A\,\left(\epsilon_{ab}\epsilon_{dc} - \epsilon_{d[a}\epsilon_{b]c}  \right)
\end{equation}
where $R^\perp_{dcab}:= h^p_a h^q_b h^n_c h^m_d R_{mnpq}$.

Due to the symmetries of the Riemann tensor and the low dimensionality, we also have  that
\begin{equation} \label{e11}
{\cal R}_{dcab} = {\cal R}\,h_{d[a} h_{b]c} = \frac{{\cal R}}{2}\,\epsilon_{dc}\epsilon_{ab}
\end{equation}
and a similar expression for $R^\perp_{dcab}$. Hence, (\ref{e10}) implies that
\begin{equation} \label{e12}
h^{da} h^{bc}R_{dcab}  = {\cal R} -\frac32\, \theta^A \theta_A \qquad {\rm and} \qquad
R^\perp_{dcab} = \left(  {\cal R} -\frac32\, \theta^A \theta_A \right) \, h_{d[a} h_{b]c}
\end{equation}

To derive the remaining components of $R_{cdab}$, namely those having some vertical indices, we use that, since $X_D^b$ is a Killing vector \cite{Stephani04},
$$R_{Dcab} := X^d_D R_{dcab} = \nabla_c \nabla_a \xi_{Db}$$
and, after some algebra we obtain:
\begin{equation} \label{e13}
R^\perp_{Dcab} = \frac12\,\left(D_c\theta_D + \frac12\,\theta^E \lambda_{DE|c}\right)\,\epsilon_{ab}
\end{equation}

\begin{equation} \label{e14}
R^\perp_{DcAb} = -\frac12\,\left(D_b\lambda_{AD|c} - \frac12\,\lambda^{TE}\lambda_{AE|c}\lambda_{TD|b} - \frac12\, \theta_A \theta_D h_{bc}\right)
\end{equation}

\begin{equation} \label{e15}
R^\perp_{DCAb} = -\frac12\,\theta_{[C}\lambda_{D]A|d} \epsilon^d_{\;b} \qquad {\rm and} \qquad
R_{DCAB} = - \frac12\,h^{bc}\lambda_{C[B|c}\lambda_{A]D|b}
\end{equation}

The Riemann tensor for $\eta_{ab}$ can be reconstructed from these components according to:
\begin{eqnarray}
\hspace*{-1em}R_{dcab} \hspace*{-.5em}& =& \hspace*{-.5em}R_1 \epsilon_{dc}\epsilon_{ab} + R_2 \Omega_{dc}\Omega_{ab} + 2 P_{A[b}\xi^A_{a]}\epsilon_{dc} + 2 P_{A[c}\xi^A_{d]}\epsilon_{ab} + 2 Q_{A[b}\xi^A_{a]}\Omega_{dc} + 2 Q_{A[c}\xi^A_{d]}\Omega_{ab}  \nonumber \\ \label{e16}
 & &  + 4 \xi^D_{[d} P_{Dc]A[b}\xi^A_{a]} + \frac12\,R_3 \left(2 \epsilon_{dc}\Omega_{ab} - \epsilon_{da}\Omega_{bc} - \epsilon_{db}\Omega_{ca}
 + 2\Omega_{dc}\epsilon_{ab} - \Omega_{da}\epsilon_{bc} - \Omega_{db}\epsilon_{ca}  \right)
\end{eqnarray}
where $\Omega_{ab}$ and $\epsilon_{dc}$ are defined in (\ref{a1}) and (\ref{a2})  (see Appendix A) and
\begin{eqnarray}
\label{e17a}
R_1 & = & \frac14\,R_{dcab}\epsilon^{dc} \epsilon^{ab} = \frac12\,\left( {\cal R} -\frac32\, \theta^A \theta_A \right) \\
\label{e17b}
R_2 & = & R_{dcab}\Omega^{dc} \Omega^{ab} = - \frac1{\tau^2}\,\left(\lambda_{11|c} \lambda_{22|b} - \lambda_{12|c} \lambda_{12|b} \right)\,h^{bc}   \\
\label{e17c}
R_3 & = & \frac12\,R_{dcab} \Omega^{dc} \epsilon^{ab} = \frac1{2\tau}\,\epsilon^{bc}\lambda^{TE}\lambda_{1E|c} \lambda_{2T|b}   \\
\label{e17d}
P_{Ab} & = &  \frac12\,\left(R_{dcAb}\epsilon^{dc} \right)^\perp = \frac12\,\left(D_b \theta_A + \frac12\,\theta^T \lambda_{TA|b}\right) \\
\label{e17e}
Q_{Ab}& = &  \left(R_{dcAb}\Omega^{dc} \right)^\perp = - \frac1{\tau}\,\theta_{[1} \lambda_{2]A|c} \epsilon^c_{\;\;b}  \\
\label{e17f}
P_{DcAb} & = & R_{D(c|A|b)}^\perp =  -\frac12\,\left(D_b\lambda_{AD|c} - \frac12\,\lambda^{TE}\lambda_{AE|(c}\lambda_{TD|b)} - \frac12\, \theta_A \theta_D h_{bc}\right)
\end{eqnarray}
It is trivial to see that in expression (\ref{e16}) the first Bianchi identity is componentwise satisfied.

Equations, (\ref{e17a}-\ref{e17f}) are relations between the quotient metric and the covariant kinematical invariants of the Killing fields, on the one hand, and the ambient metric on the other. They must also be taken as equations to solve in the so called `reconstruction problem' (see next section).

\subsection{The reconstruction problem \label{S4}}
It consists in reconstructing an ambient metric $\eta_{ab}$ from a given quotient metric $h_{ab}$ provided that $\eta_{ab}$ admits two Killing vectors $X_A^b$, $A=1,2$. It is particularly interesting the case in which the final ambient metric is required to have some prescribed geometric properties, e. g. being flat, which is the case she shall ultimately be interested in.

It is easy to prove that giving a metric $\eta_{ab}$ on $\mathcal{M}$ is equivalent to providing:
\begin{list}
{(a.\roman{llista})}{\usecounter{llista}}
\item two covectors $\xi_{Aa} \in \Lambda^1 {\mathcal{M}}$ such that $\mathcal{L}_{X_B}\xi_{Aa} = 0 $ and that $\lambda_{AB}:=\xi_{Aa} X_B^a$ is a non-degenerate matrix, and
\item the quotient metric on $\mathcal{S}$. (The signatures of both $h_{ab}$ and $\lambda_{AB}$ must be chosen so that the signature of $\eta_{ab}$ is $(+3,-1)$.)
\end{list}
On their turn these conditions are equivalent to giving:
\begin{list}
{(b.\roman{llista})}{\usecounter{llista}}
\item two covectors $\xi^A_{a} \in \Lambda^1 {\mathcal{M}}$ such that $\mathcal{L}_{X_B}\xi_{Aa} = 0 $ and that
$\xi^A_{a} X_B^a = \delta^B_A$,
\item a 2-squared symmetric non-degenerate matrix $\lambda_{AB} \in \Lambda^0 {\mathcal{S}}$ and
\item the quotient metric on $\mathcal{S}$.
\end{list}


\subsection{Reconstructing a flat metric with two prescribed Killing vectors \label{SS2.4}}
Assume now that we want the ambient metric to be flat. Are there any further restrictions on $h_{ab}$, $\xi^A_a$ and $\lambda_{AB}$ that are derived from the flatness of $\eta_{ab}$?

As $\eta_{ab} X^a_B = \xi_{Ba}$ and $X_A^a$, $A=1,2$, are Killing vectors, the results derived in section \ref{SS3.1} apply. Therefore, $\mathcal{L}_{X_B} \xi^A_{a} =0$,  $\mathcal{L}_{X_B} \lambda_{AB} =0$ and equations similar to (\ref{e8}) do hold. Thus, although $\lambda_{AB}\in\Lambda^0(\mathcal{S})$, $\xi_{Aa}$ and $\xi^A_a$ are not covectors on $\mathcal{S}$ because they are not orthogonal to $X_B^b$. Let us assume however that we are given two covectors $\overline{\xi}^A_{a}\in\Lambda^1(\mathcal{M})$ such that $\overline{\xi}_{Ab}X_B^b=\delta^A_{B}$, $A,B=1,2$ and that $\mathcal{L}_{X_B}\overline{\xi}^A_{a}= 0$. Then  $\xi^A_{a}$ can be written as
\begin{equation} \label{e20b}
\xi^A_{a} =\overline{\xi}^A_{a} + \kappa^A_a \,, \qquad {\rm with}  \qquad  \kappa^A_a\in \Lambda^1(\mathcal{S})
\end{equation}
We shall call $\kappa^A_a$ the {\em shift covectors}, differentiating and taking (\ref{e8}) into account, we arrive at:
\begin{equation} \label{e20c}
\theta^A \epsilon = d\overline{\xi}^A + d\kappa^A
\end{equation}
Bearing  this result in mind, the expressions (\ref{e17a}-\ref{e17f}) imply a second order partial differential system on the variables $\lambda_{AB}$, $\kappa^A_a$ i $h_{ab}$, namely
\begin{equation}  \label{e20d}
R_1 = R_2 = R_3 \,,\qquad P_{Ab} = Q_{Ab}=0 \,, \qquad P_{DcAb} = 0 \,,
\end{equation}
which has to be solved on $\mathcal{S}$ and the solutions are to be used as the data (b.i) to (b.iii) necessary to reconstruct $\eta_{ab}$.

In Appendix B we prove that equations (\ref{e20d}) imply that $h_{ab}$, $\xi^A_a$ and $\lambda_{AB}$ are constrained by the following conditions
\begin{itemize}
\item The horizontal metric $h_{ab}$ must be either (i) flat or, if not, (ii) its Ricci scalar must satisfy
\begin{equation}  \label{Mb5a}
D_{bc} \mathcal{R}^{-1/4}  +  \frac{1}{6}\,\mathcal{R}^{3/4}\, h_{bc} = 0
\end{equation}
\item In case (i), take $\theta_A=0$ and
\begin{list}
{{\bf i.\alph{llista}.-}}{\usecounter{llista}}
\item either take [equation (\ref{R141a}), $q=0$ ]
\begin{equation} \label{R141aa}
\lambda_{AB} = \frac{\tau_0}{\sqrt{2}}\,\left[  \hat\lambda_{AB} +  F\,\hat{q}_{AB} \right]
\qquad {\rm with}  \qquad {\rm i.e.} \qquad D_{ab}F =0
\end{equation}
where $\hat\lambda_{AB}$ and $\hat q_{AB}$ are constant matrices fulfilling
$ \hat q_{AB}\hat\lambda^{AB} =  \hat q_{AB}\hat \lambda^{BC}\hat q_{CD}= 0 \,$, and  $\det(\hat\lambda_{AB})=-1  $.
\item or take [equation (\ref{R142}), $q=1$ ]
\begin{equation} \label{R142aa}
\lambda_{AB} = \frac1{\sqrt{2}}\,\left( - \Phi_-^2 \hat m_A \hat m_A + \Phi_+^2 \hat n_A \hat n_A \right) \qquad {\rm with} \qquad \mathcal{F}_{bc}^\pm:= D_{bc} \Phi_\pm =0
\end{equation}
with $ \hat m_2 \hat n_1 - \hat m_1 \hat n_2 = 1$ and $\Phi_\pm$ fulfilling (\ref{R142a}).
\end{list}
\item In case (ii) choose two constants $\alpha \neq 0$ and $C\neq 0$ and take
$$ \tau = \left(-\,\frac{3\alpha C^2}{\mathcal{R}} \right)^{1/4} \,, \qquad \theta = \frac{C}{\tau} = \left(-\,\frac{ C^2 \mathcal{R}}{3\alpha} \right)^{1/4} \,,$$
then choose $k_A$ and $\hat{\lambda}^{AB}$ such that $\hat{\lambda}^{AB} k_A k_B = \alpha$ and take
$$ \theta_A = k_A \theta\,, \qquad \theta^A = -\,\frac{2C}{\tau^3}\,\hat{\lambda}^{AB} k_B \,, \qquad\lambda_{AB}= k_A k_B \lambda + \hat{\lambda}_{AB} \quad {\rm with} \quad \lambda :=-\frac1{\alpha}\,\left( \frac{\tau^2}{2} + \delta_0\right) $$
\end{itemize}

In both cases we still have  to determine $\xi^A \in \Lambda^1\mathcal{M}$. To this aim, we first choose two 1-forms $\overline\xi^A \in \Lambda^1\mathcal{M}$ such that $\mathcal{L}_{X_B} \overline\xi^A = 0$ and $X_B^a  \overline\xi^A_a= \delta^A_B$; it is obvious that $d\overline\xi^A\in \Lambda^2\mathcal{S}$. The shift covector, $\kappa^A :=\xi^A - \overline\xi^A \in \Lambda^1\mathcal{S}$, can be determined by solving
\begin{equation}   \label{MR16}
d \kappa^A = \theta^A \epsilon - d\overline\xi^A
\end{equation}
which follows from (\ref{e8}) and is always integrable due to the fact that dim$\,\mathcal{S}= 2$.

\section{Flat deformation \label{S5}}
The central result of the present paper is the following theorem.

\begin{theorem}   \label{t1}
Let $g_{ab}$ be a Lorentzian metric admitting two commuting Killing vector fields, $X_A^a$, $A=1,2$. Then, there exist two functions $a$, $b$ and an elliptic 2-dimensional projector $H^a_b$ such that the deformed metric
\begin{equation} \label{e23}
\eta_{ab}:= a\,g_{ab} + b\,H_{ab} \,,\qquad {\rm where} \quad  H_{ab}:=g_{ac} H^c_b \,,
\end{equation}
is flat and admits $X_A^a$, $A=1,2$, as Killing vector fields with vanishing twists.
\end{theorem}

The proof spreads all over the present section but we first need to review some useful results.
\begin{lemma}  \label{P8}
Let $X^a$ be a Killing vector for $g_{ab}$ and let $\eta_{ab}$ be defined by (\ref{e23}) with $b\neq 0$, then:
 \begin{equation} \label{e23a}
{\mathcal{L}}_X\eta_{ab} = 0 \qquad \Leftrightarrow \qquad {\mathcal{L}}_X a = {\mathcal{L}}_X  b = 0 \qquad {\rm and} \qquad {\mathcal{L}}_X H_{ab}=0
\end{equation}
\end{lemma}
See \cite{CarotLlosa08} for a proof. \hfill $\Box$.

With respect to the Killing vectors, the given metric $g_{ab}$ splits into its horizontal and vertical parts as follows:
 \begin{equation} \label{p1a}
 g_{ab}= \overline{h}_{ab}+ \overline\lambda_{AB}\overline\xi^A_{a} \overline\xi^B_b
 \end{equation}
where $\overline{h}_{ab}$ is the quotient metric and $\overline\lambda_{AB}\overline\xi^A_{a} \overline\xi^B_b$ is the metric on the orbits, with
\begin{equation} \label{e22a}
\overline\xi_{Aa}:=g_{ab}X_A^b\,, \quad \overline\lambda_{AB}:=g_{ab}X_A^a X_B^b \,, \quad \overline\xi^A_{a}:=\overline\lambda^{AB}\overline\xi_{Ba} \quad {\rm and} \quad \overline\lambda^{AB} \overline\lambda_{BC}=\delta^A_C
\end{equation}
On its turn, the sought-after deformed metric $\eta_{ab}$ may also be split into its horizontal and vertical parts 
 \begin{equation} \label{p1}
 \eta_{ab} = h_{ab} +\lambda_{AB} \xi^A_a \xi^B_b
 \end{equation}
with $\xi^A_a$ and $\lambda_{AB}$ defined as in (\ref{e1}).

Since $X_A^a$ are commuting Killing vectors for both $g_{ab}$ and $\eta_{ab}$, we have that $\mathcal{L}_{X_A} \xi^B_a=\mathcal{L}_{X_A} \overline\xi^B_a= 0$ and, as a consequence, we can introduce the shift covectors
 \begin{equation} \label{p2}
 \kappa^A_a := \xi^A_a -\overline\xi^A_a \in \Lambda^1\mathcal{S}
 \end{equation}

As commented above ---conditions (b.i) to (b.iii) in subsection \ref{S4}--- to determine $\eta_{ab}$ is equivalent to finding an elliptic horizontal metric $h_{ab}$, the hyperbolic matrix $\lambda_{AB}$, $A, B = 1,2$, and two covectors $\xi^A_a$ such that $\xi^A_a X_B^a =\delta_B^A$.

\subsection{The unknowns \label{SS2.1}}
To relate these objects with the unknowns $a$, $b$ and $H_{ab}$ in the deformation law (\ref{e23}), we shall take into account that the elliptic 2-dimensional projector and can be written as:
\begin{equation} \label{E1}
H_{ab} = m_a m_b + n_a n_b \,, \qquad {\rm with} \qquad m_a m_b g^{ab}=  n_a n_b g^{ab}=  1 \qquad {\rm and } \qquad
m_a n_b g^{ab}= 0
\end{equation}
The covectors $m_a$ and $n_a$ can then be written in terms of their respective vertical and horizontal components (relatively to the vectors $X^a_B$ and the metric $g_{ab}$):
\begin{equation} \label{E2}
m_b =m_A \overline{\xi}^A_b + \mu_b \,, \qquad \qquad n_b =n_A \overline{\xi}^A_b + \nu_b
\end{equation}
where $m_A:=m_a X^a_A$, $n_A:=n_a X^a_A$ and $\mu_a X^a_A = \nu_a X^a_A = 0$. In Appendix C ---equations (\ref{A2}), (\ref{A3}) and proposition \ref{Pz2}--- we prove that the covectors $m_a$ and $n_a$ may be chosen so that:
\begin{list}
{(\alph{llista})}{\usecounter{llista}}
\item $\mathcal{L}_{X_B} m_A = \mathcal{L}_{X_B} n_A = 0\,$, hence $m_A,\, n_A \in \Lambda^0\mathcal{S}$;
\item $\,\mathcal{L}_{X_B} \mu_a = \mathcal{L}_{X_B} \nu_a = 0$, hence $\mu_a,\, \nu_a \in \Lambda^1\mathcal{S}$ and
\item if we put $x:=m_A m_B \overline{\lambda}^{AB}$ and $y:=n_A n_B \overline{\lambda}^{AB}$, then\footnote{We confine ourselves to the generic case $x y \neq 0$. The fully degenerate case $x=y=0$ is studied in detail in section \ref{S6}}
\begin{equation} \label{E3}
\mu_a\mu_b\overline{h}^{ab}= 1-x \,,\qquad \nu_a\nu_b\overline{h}^{ab}= 1-y \qquad {\rm and} \qquad \mu_a\nu_b\overline{h}^{ab}= 0
\end{equation}

\end{list}
Also in Appendix C it is shown ---equation (\ref{A7})--- that the shift covectors are
\begin{equation} \label{E4}
\kappa_a^A = b  \,\left( m^A \mu_a + n^A \nu_a\right)  \,, \qquad {\rm with} \qquad
m^A:= \lambda^{AB} m_B \,, \qquad n^A:= \lambda^{AB} n_B
\end{equation}

Following the guidelines advanced in Appendix B 
to ensure that the components (\ref{e20d}) of the Riemann tensor for $\eta_{ab}$ do vanish we choose
\begin{list}
{(\alph{llista})}{\usecounter{llista}}
\item \underline{$\theta^A=0$, $A=1,2$} and, including (\ref{p2}), the exterior derivatives of the shift covectors are
\begin{equation} \label{F1}
\left(d\kappa^A\right)_{ab} = -\overline{\theta}^A \overline{\epsilon}_{ab}
\end{equation}
$\overline\theta^A$ and $\overline{\epsilon}_{ab}$ respectively being the twist of $\overline\xi^A_b $ and the volume tensor for $\overline{h}_{ab}$.
\item \underline{The matrix $\lambda_{AB}$} given by the expression (\ref{R142aa}) 
\begin{equation} \label{F2}
\lambda_{AB} = \frac1{\sqrt{2}}\,\left( - \Phi_-^2 \hat m_A \hat m_B + \Phi_+^2 \hat n_A \hat n_B \right) 
\qquad {\rm with} \qquad D_{bc} \Phi_\pm =0 \,,
\end{equation}
where $\hat m_A$ and $\hat n_A$ are constants such that $ \hat m_2 \hat n_1 - \hat m_1 \hat n_2 = 1$. 

Furthermore, using (\ref{E4}) and  (\ref{A4}-\ref{A6}) in Appendix C, we obtain
\begin{equation} \label{F6a}
H_{ab} = H_{AB} \overline\xi^A_a \overline\xi^B_b + \mu_a \mu_b + \nu_a \nu_b + 2 \overline\xi^A_{(a} \kappa^B_{b)} \lambda_{AB}
\end{equation}
where
\begin{equation} \label{F6b}
 b H_{AB}= \lambda_{AB} - a \overline\lambda_{AB} = \frac1{\sqrt{2}}\,\left( - \Phi_-^2 \hat m_A \hat m_B + \Phi_+^2 \hat n_A \hat n_B \right) - a \overline\lambda_{AB}  
\end{equation}
gives $H_{AB}$ in terms of $\Phi_\pm$, $a$ and $b$.
\item \underline{A flat horizontal metric $h_{ab}$} which, according to equation (\ref{A9}) in Appendix C, is
\begin{equation}  \label{F7}
h_{ab} = a (a+b) \tilde{h}_{ab} \,, \qquad {\rm with} \qquad
\tilde{h}_{ab}:= z_1 \hat\mu_a \hat\mu_b + z_2 \hat\nu_a \hat\nu_b
\end{equation}
where $z_1$ and $z_2$ are given by (\ref{F6}), $\{\hat\mu_a,\,\hat\nu_a\}$ is an $\overline h$-orthonormal base and
\begin{equation}  \label{F7a}
\hat\mu_a :=(1-x)^{-1/2} \mu_a  \qquad {\rm and} \qquad  \hat\nu_a :=(1-y)^{-1/2} \nu_a
\end{equation}
The Ricci scalar is 
\begin{equation}  \label{F7aa}
\mathcal{R} = 0
\end{equation}
\end{list}

Finally, from equation (\ref{A7a}) in Appendix C, we have that
\begin{equation} \label{F6c}
\left. \begin{array}{l}
       \displaystyle{m_A =  2^{-1/4} \sqrt{-x z_1} \,\left( \Phi_- \cosh\zeta\, \hat{m}_A +\Phi_+ \sinh\zeta \,\hat{n}_A  \right)    } \\
 \displaystyle{n_A = 2^{-1/4} \sqrt{y z_2} \,\left( \Phi_- \sinh\zeta \,\hat{m}_A + \Phi_+ \cosh\zeta \,\hat{n}_A  \right)   }
       \end{array}  \right\}
\end{equation}  
where $z_{1,2}$, $x$, $y$ and $\zeta$ are functions of $\Phi_\pm$, $a$ and $b$ [see equations (\ref{F6}), (\ref{F6ab}) and (\ref{F6ac})]. Substituting these in (\ref{E4}) yields an expression for the shift covectors in terms of the new unknowns $\Phi_\pm$, $a$, $b$ and the orthonormal base $\{\hat\mu_a,\,\hat\nu_a\}$.

\subsection{The equations \label{SS2.2}}
Equation (\ref{F6a}), combined with equations   (\ref{E4}), (\ref{F6}), (\ref{F6b}) and (\ref{F6c}), give  $H_{ab}$ in terms of $a$, $b$, $\Phi_\pm$ and the $\overline{h}$-orthonormal base $\{\hat\mu_a,\,\hat\nu_a\}$. The unknowns $\Phi_\pm$ are governed by the equations (\ref{F2}), while the remaining variables are ruled by equations (\ref{F1}) and (\ref{F7aa}).

Let us first examine the equations (\ref{F1}) which, on account of (\ref{E4}), can be takien as differential equations on $\mu_a$ and $\nu_a$. Since $\{\hat\mu_a,\,\hat\nu_a\}$ is an $\overline{h}$-orthonormal base of $T\mathcal{S}$, we can write
\begin{equation} \label{W1}
\overline{D}_a \hat\mu_b - \omega_a \hat\nu_b = 0 \qquad {\rm with} \qquad \hat\nu_b=\hat\mu_a\overline\epsilon^a_{\;\,b}\,,
\qquad \omega_a \in \Lambda^1\mathcal{S} \,,
\end{equation}
where $\overline{D}$ is the Levi-Civita connection for $\overline{h}_{ab}$; therefore
\begin{equation} \label{W1a}
d\hat\mu = \omega \wedge\hat\nu  \qquad {\rm and} \qquad d\hat\nu = -\omega \wedge\hat\mu
\end{equation}

Let $\{ {\hat\alpha}^3_a,\,{\hat\alpha}^4_a\}$ be a given $\overline{h}$-orthonormal base and let $\{\hat{\bm e}_3^a,\, \hat{\bm e}_4^a\}$ be the dual base. We then have that it exists $\psi\in\Lambda^0\mathcal{S}$, such that
\begin{equation} \label{W1b}
\hat\mu_a = \cos\psi\,\hat\alpha^3_a  -\sin\psi\, \hat\alpha^4_a\,, \qquad  \hat\nu_a = \sin\psi\,\hat\alpha^3_a  +\cos\psi\, \hat\alpha^4_a
\end{equation}
From this and (\ref{W1}) we readily obtain that $ (\psi_a + \omega_a)\hat\nu_b + \hat{\mu}_c \overline\gamma_{ab}^c =0 $ and write:
\begin{equation} \label{W1c}
\mathcal{E}_{a}:= \hat{\bm e}_a\psi + \omega_b \hat{\bm e}_a^b + \hat{\mu}_c \overline\gamma_{ab}^c \hat{\nu}^b =0 \,,
\end{equation}
where $\overline\gamma_{a\;b}^{\;\,c}$ are the connection coefficients for $\overline{D}$.

Then, combining (\ref{E4}), (\ref{F1}),(\ref{F7a}) and (\ref{W1a}), we arrive at
$$ d\left(b m^A\sqrt{1-x} \right) \wedge\hat\mu + d\left(b n^A\sqrt{1-y} \right) \wedge\hat\nu + \overline\theta^A \overline\epsilon=
\left(b m^A\sqrt{1-x}\,\hat\nu - b n^A\sqrt{1-y}\,\hat\mu \right) \wedge \omega   $$
and, writing $\omega = \Omega^1 \hat\mu + \Omega^2 \hat\nu$, we finally obtain
\begin{equation} \label{W3}
b m^A\sqrt{1-x}\,\Omega^1 + b n^A\sqrt{1-y}\,\Omega^2 = \rho^A  \,, \qquad A=1,\,2
\end{equation}
where
\begin{equation} \label{W4}
\rho^A := - \overline\theta^A - \hat\mu^b \,\left(b n^A\sqrt{1-y} \right)_{|b} + \hat\nu^b \,\left(b m^A\sqrt{1-x} \right)_{|b}
\end{equation}
with $\hat\mu^b:= \overline{h}^{bc}\hat\mu_c$ and so on.

Equations (\ref{W3}) can be solved providing expressions for $\Omega^1$ and $\Omega^2$  in terms of $a$, $b$, $t$ and $f$:
\begin{equation} \label{W4a}
\Omega^1 = \frac{\tau^2}{b\overline\tau\,\sqrt{-2xy(1-x)}}\,\left(\rho^1 n^2 - \rho^2 n^1 \right) \qquad {\rm and} \qquad
\Omega^2 = \frac{\tau^2}{b\overline\tau\,\sqrt{-2xy(1-y)}}\,\left(\rho^2 m^1 - \rho^1 m^2 \right)
\end{equation}
where (\ref{A3A}) has been included.
In their turn, the components of $\omega_a$ in the given orthonormal base are:
\begin{equation} \label{W4aa}
\omega_a \hat{\bm e}_3^a = \Omega^1 \cos\psi + \Omega^2 \sin\psi \qquad {\rm and} \qquad
\omega_a \hat{\bm e}_4^a = -\Omega^1 \sin\psi + \Omega^2 \cos\psi
\end{equation}

Once $\omega_a$ is known, we may substitute it in equations (\ref{W1c}) which then become a partial differential system on $\psi_a$. As too many derivatives of the unknowns are specified, an integrability condition is implied which, since $\dim\,\mathcal{S}=2$, reads
$ \overline{D}_a \omega_b - \overline{D}_b \omega_a = -\frac{\overline{\mathcal{R}}}{2}\, \overline\epsilon_{ab} $ or
\begin{equation} \label{W5}
\mathcal{W}:= d \omega +\frac12\,\overline{\mathcal{R}} \overline\epsilon = 0
\end{equation}
Now, taking (\ref{W4a}) and (\ref{W4aa}) into account, the above is a second order partial differential equation on the unknowns $a$, $b$, $\Phi_\pm$ and $\psi$.

We are thus led to solving the partial differential system constituted by equations (\ref{R142aa}), (\ref{F7aa}), (\ref{W1c}), (\ref{W5})  and  the flatness of $h_{ab}$:
\begin{equation} \label{W7}
\qquad R_{2}=0 \,, \qquad \mathcal{F}^\pm_{ab}=0  \,, \qquad \mathcal{E}_{a}=0 \,, \qquad \mathcal{W}=0 \,, \qquad \mathcal{R}=0
\end{equation}  

The number of equations exceeding by far the number of unknowns, we shall deal much in the same way as it is usually done with Einstein equations: considering  a certain subset of distinguished equations as the {\em reduced PDS}, and treating the remaining  equations as {\em constraints}; the existence of solutions will then be studied in terms of a Cauchy problem.

We choose a hypersurface $\Sigma$ in $\mathcal{S}$ which will act as a Cauchy hypersurface for the partial differential system ($\Sigma$ is actually a curve because ${\cal S}$ has two dimensions), and take Gaussian $\overline h$-normal coordinates $(y^3,y^4)$ in a neighbourhood of $\Sigma$, so that $y^4=0$ on $\Sigma$ and
\begin{equation} \label{W8}
 \overline h_{34}=\overline h^{34}= 0 \,, \qquad \overline h_{44}=\overline h^{44}= 1 \,, \qquad \overline h_{33}= U  \qquad  {\rm and}
 \qquad    \overline h^{33}= U^{-1}
\end{equation}
Thus, the above mentioned $\overline{h}$-orthonormal may be taken to be
$$\hat{\bm e}_3^a = \delta^a_3 \,\frac1{\sqrt{U}}\,, \qquad
\hat{\bm e}_4^a = \delta^a_4  \qquad {\rm and} \qquad
{\hat\alpha}^3_a = \delta^3_a \,\sqrt{U}\,, \qquad
{\hat\alpha}^4_a = \delta^4_a \,.  $$

We can now consider (\ref{W7}) as a system of differential equations in the five unknowns: $\Phi_\pm$, $a$, $b$  and $\psi$, and separate:
\begin{description}
\item[the reduced system,] namely
\begin{equation} \label{W9}
\mathcal{F}^\pm_{44}=0 \,, \qquad \mathcal{E}_{4}=0 \,, \qquad
\mathcal{W}=0 \qquad {\rm and} \qquad \mathcal{R}=0 \,, \quad {\rm and}
\end{equation}
\item[the constraints:]
\begin{equation} \label{W10}
\mathcal{F}^\pm_{3a}=0 \,, \qquad R_2=0 \,, \qquad  {\rm and} \qquad
\mathcal{E}_{3}=0 \,,
\end{equation}
($a= 3,4$)to be satisfied by the Cauchy data on $\Sigma$.
\end{description}

\begin{proposition}   \label{pr.4}
Any solution of the reduced system (\ref{W9}) fulfilling the constraints (\ref{W10}) on the Cauchy hypersurface $\Sigma$ also fulfills the constraints in an open neighbourhood of $\Sigma$.
\end{proposition}

\paragraph{Proof:} 
By differentiating  (\ref{R142aa}) and (\ref{W1c}), we easily obtain (recall that we have chosen $q=1$)
\begin{equation}  \label{W12a}
 D_{[c}\mathcal{F}^\pm_{a]b} \equiv  -  \frac{\mathcal{R}}{2} D_d \Phi_\pm \epsilon^{d}_{\;b} \epsilon_{ca} \,, \qquad
D_a R_2 \equiv  \frac12\,h^{bc}\left(\mathcal{F}^+_{ab} D_c \Phi_- + \mathcal{F}^-_{ab} D_c \Phi_+\right) \qquad {\rm and} \qquad 
\overline{D}_{[a} \mathcal{E}_{b]} \equiv  \frac{\mathcal{W}}{2}\,\overline\epsilon_{ab}  
\end{equation}
and, particularly, since we are dealing with a solution of the reduced system (\ref{W9}), we obtain for the constraints:
$$  D_4 R_2 = \frac12\,U^{-1}\,\left(\mathcal{F}^+_{43} D_3\Phi_- + \mathcal{F}^-_{43} D_3\Phi_+\right)
 \,, \qquad D_{4}\mathcal{F}^\pm_{3a} = D_{3}\mathcal{F}^\pm_{43}\delta^3_a  
\qquad {\rm and} \qquad  \overline{D}_4 \mathcal{E}_{3} = 0\,, $$
$a=3,4 $, which is a linear, homogeneous, partial differential system to be fulfilled by the constraints, whence it follows that the vanishing of the constraints on $\Sigma$ propagates to an open neighbourhood of $\Sigma$.  \hfill $\Box$ 

\subsection{The reduced system. Characteristic determinant}
To decide whether $\Sigma$ is a non-characteristic hypersurface for the reduced system (\ref{W9}) we must study its characteristic determinant \cite{John}. To this end we must consider only the principal part of its equations, i. e. the part containing the highest order derivatives of the unknowns. Particularly, $\partial^2_4 \Phi_\pm$, $\partial^2_4 a$, $\partial^2_4 b$ and $\partial_4 \psi$, and we easily obtain that:
$$  \mathcal{F}^\pm_{44} \cong \partial^2_4 \Phi_\pm \,, \qquad
\mathcal{E}_{4} \cong \partial_4 \psi $$
where $\cong$ means ``equal apart from non-principal terms''.

The principal parts of the remaining two equations are not so simple; they look like
\begin{eqnarray*}
\mathcal{W} & = & \mathcal{W}^{+} \partial^2_4 \Phi_+ + \mathcal{W}^{-} \partial^2_4 \Phi_- + \mathcal{W}^{(\psi)} \partial_4 \psi + \mathcal{W}^{(a)} \partial^2_4 a + \mathcal{W}^{(b)} \partial^2_4 b  \\
\mathcal{R} & = & \mathcal{R}^{+} \partial^2_4 \Phi_+ + \mathcal{R}^{-} \partial^2_4 \Phi_- + \mathcal{R}^{(\psi)} \partial_4 \psi + \mathcal{R}^{(a)} \partial^2_4 a + \mathcal{R}^{(b)} \partial^2_4 b
\end{eqnarray*}
It easily follows that the characteristic determinant of the reduced system is
$ \chi = \mathcal{W}^{(a)}  \mathcal{R}^{(b)} - \mathcal{W}^{(b)}  \mathcal{R}^{(a)} $
and we do not need to calculate explicitly all the coefficients in the principal part of $\mathcal{W}$ and $\mathcal{R}$.
A detailed, heavy-going calculation yields
\begin{eqnarray}  
\chi & =& -\,\frac{z_1 + z_2 + (z_1 - z_2) \cos 2\psi}{4 a^2(a+b)^2 z_1 z_2}\,\left( \frac1{f(a+b)}-\frac1{f(a)}  \right. \\
 & & \left.\frac12\, \left[z_1 + z_2 - (z_1 - z_2) \cos 2\psi\right] \,\left[\frac{a+b}{f(a+b)}-\frac{a}{f(a)}  \right] 
 \label{W13}
\right)
\end{eqnarray}
where $f(a):= 1 - a (z_1+z_2) + a^2 z_1 z_2$ and, according with (\ref{F6}), 
\begin{equation}  \label{W14}
z_1 + z_2 = \overline\lambda_{AB} \lambda^{AB}\,, \qquad z_1 z_2 = \overline\tau^2/\tau^2 \qquad {\rm and} \qquad 
z_1 - z_2 = \sqrt{(\overline\lambda_{AB} \lambda^{AB})^2 - 4 \overline\tau^2/\tau^2}
\end{equation}
Then, in order that $\Sigma$ is a non-characteristic hypersurface, Cauchy data must be chosen so that $\chi \neq 0$.

\subsection{The constraints}
The Cauchy data, namely $\Phi_\pm$, $\dot\Phi_\pm:=\partial_4 \Phi_\pm$, $a$, $b$, $\dot a$, $\dot b$ and $\psi$ on $\Sigma$, must be chosen so that $\chi \neq 0$ and the constraints (\ref{W10}) are fulfilled. $\Sigma$ is a curve and the coordinate $u:=y^3$ acts as a curve parameter; the constraints can thus be written as
\begin{equation}  \label{equates}
\left. \begin{array}{l}
\displaystyle{\mathcal{F}^\pm_{34} :=  \frac{d\dot \Phi_\pm}{du} - \Gamma^{\;\,c}_{3\;4} D_c\Phi_\pm = 0 } \\[1ex]
\displaystyle{\mathcal{F}^\pm_{33} :=  \frac{d^2 \Phi_\pm}{du^2} - \Gamma^{\;\,c}_{3\;3} D_c\Phi_\pm = 0 } \\[1ex]
\displaystyle{\mathcal{E}_{3} :=  \frac{d \psi}{du} - \overline\Gamma^{\;\,4}_{3\;3} \,\frac1{\sqrt{U}} + \omega_3 = 0 }
\end{array} \right\}
\end{equation}
$(c=3,4)$, and we must replace $D_3\Phi_\pm$ by $\displaystyle{\frac{d \Phi_\pm}{du}}$, $D_4\Phi_\pm$ by $\dot\Phi_\pm$ and so on. $\Gamma^c_{ed}$ and $\overline\Gamma^c_{ed}$ are the Christoffel symbols for the connections $D$ and $\overline{D}$, respectively, and they depend on $\Phi_\pm$, $a$, $b$, their first order derivatives and $\psi$.

We can therefore prescribe arbitrary values for $a$, $\dot a$, $b$ and $\dot b$ on $\Sigma$, because there is no constraint on them, and then substitute them into (\ref{equates}) which can be taken as an ordinary differential system on the remaining Cauchy data:  $\Phi_\pm$, $\dot\Phi_\pm$ and $\psi$ on the curve $\Sigma$. This system admits a solution for any given initial data $\Phi_\pm(x_0)$, $D_3\Phi_\pm(x_0)$ and $\psi(x_0)$, for a given point $x_0\in\Sigma$.

As for the remaining constraint, $\displaystyle{R_2:= \frac12\, h^{bc} D_b\Phi_+ D_c \Phi_- = 0}$, it acts merely as a condition on the initial data $D_c\Phi_\pm(x_0)$. (Notice that equation (\ref{W12a}) implies that, if $R_2(x_0)=0$ then $R_2=0$ in a neighbourhood of $x_0$.)

\subsection{Summary of the proof}
We now show how to construct the deformed metric $\eta_{ab}$ from a solution to the above Cauchy problem.

\begin{list}
{(\arabic{llista})}{\usecounter{llista}}
\item Take an analytic curve $\Sigma_0 \subset \mathcal{S}$ and choose a point $x_0\in \Sigma_0$.
\item {\bf Give} $a$, $\dot a$, $b$ and $\dot b$, analytic functions on $\Sigma_0$, then
\item {\bf give}\footnote{These data must be chosen so that the conditions (\ref{W14}) are fulfilled} $\psi(x_0)$,
$\Phi_\pm(x_0)$ and  $D_c\Phi_\pm(x_0)$, $(c=3,4)$.
\item Then solve the ordinary differential system (\ref{equates}) to obtain $\psi$, $\Phi_\pm$ and $\dot\Phi_\pm$  on $\Sigma$, a neighbourhood of $x_0$ in $\Sigma_0$.
\item With these $a$, $b$, $\dot a$, $\dot b$, $\Phi_\pm$, $\dot\Phi_\pm$ and $\psi$ as Cauchy data on $\Sigma$, the {\em reduced system } has an analytic solution,
$$ a, \quad b, \quad \Phi_\pm \,, \quad \psi \qquad \mbox{on a neighbourhood of} \quad \Sigma $$.
\item With $\psi$ and equation (\ref{W1b}) construct the $\overline{h}$-orthonormal basis $\hat\mu_c$, $\hat\nu_d$. 
\item {\bf Give} constant $\hat m_A$ and $\hat n_A$, $A=1,2$, such that $(\hat m_1 \hat n_2 - \hat m_2 \hat n_1 )^2= 1$
and substitute in formula (\ref{F2}) to obtain
$$ \frac1{\sqrt{2}}\,\left( - \Phi_-^2 \hat m_A \hat m_A + \Phi_+^2 \hat n_A \hat n_A \right) $$
\item Taking (\ref{p2}) and (\ref{E4}) into account, we obtain the covectors 
$$\xi^A_a = \overline\xi^A_a + b\left(m^A \hat\mu_a\,\sqrt{1-x} + n_A \hat\nu_a\,\sqrt{1-y} \right)$$ 
where $m_A$ and $n_A$ are given by (\ref{A7a}), then  
\item using (\ref{F7}) we derive the horizontal metric $h_{ab}$ and
\item we finally obtain the deformed metric $\eta_{ab}= h_{ab}+\lambda_{AB} \xi^A_a \xi^B_b$.
\end{list} 

\section{A particularly simple case \label{S6}}
We shall now consider the fully degenerate case $x=y=0$, which implies that $H_{ab}$ is a horizontal tensor and, as it is a 2-dimensional projector in a 2-dimensional space, $H_{ab}= \overline h_{ab}$. Therefore the original and deformed metrics respectively are
\begin{equation}  \label{n1}
g_{ab} = \overline h_{ab} + \overline k_{ab} \qquad {\rm and} \qquad  \eta_{ab} = \varphi \overline h_{ab} + a \overline k_{ab} 
\end{equation}
with $\overline k_{ab}:=\overline\lambda_{AB} \overline\xi^A_a \overline\xi^B_b\,$ and $\,\varphi :=a + b$. 
In this case, which we shall refer to as a {\em degenerate deformation law}, the 2-planes of the almost-product structure $g_{ab} = H_{ab} + K_{ab}$ which, by a biconformal deformation yields the flat metric $\eta_{ab}$, coincide with the almost-product structure associated to the orbits of the Killing fields.

This is indeed a non-generic case: a metric  $g_{ab}$ with two commuting Killing vectors does not, in general, admit a  {\em degenerate deformation law} yielding a flat $\eta_{ab}$. We shall here characterize the metrics $g_{ab}$ admitting a degenerate deformation.


From (\ref{n1}) we have that
\begin{equation}  \label{n2}
\xi_{Aa} = a \overline\xi_{Aa} \,, \qquad \lambda_{AB} = a \overline\lambda_{AB} \,, \qquad \lambda^{AB} = \frac1{a}\, \overline\lambda^{AB}\,, \qquad \xi^A_b =\overline\xi^A_b \qquad {\rm and}  \qquad h_{ab} = \varphi \overline h_{ab}
\end{equation}
whence it follows that
\begin{equation}  \label{n3}
\tau = a \overline\tau \,, \qquad \overline\theta^A = \varphi \theta^A  \qquad {\rm and} \qquad 
\overline\theta_A =  \theta_A \varphi/a
\end{equation}
where the fact that $\epsilon_{ab} = \varphi \overline\epsilon_{ab}$ has been included. 

Since $\eta_{ab}$ is flat and has two commuting Killing vectors (see Appendix B), only two possibilities are left:
\begin{list}
{(\alph{llista})}{\usecounter{llista}}
\item $\theta_1=\theta_2 = 0$ which, by (\ref{n3}), implies that $\overline\theta_1=\overline\theta_2 = 0$ and
\item $\theta_A= k_A \theta$, wiht $k_A=\,$ constant, which implies that $\overline\theta_A= k_A \overline\theta$.
\end{list}

Notice that $\tilde{\overline{\lambda}}_{AB}:=\overline\tau^{-1} \overline\lambda_{AB} \sqrt{2} = \tau^{-1} \lambda_{AB} \sqrt{2} = \tilde\lambda_{AB}$ and, as this $\tilde\lambda_{AB}$ corresponds to the metric $\eta_{ab}$ which is flat, the results derived in Appendix B apply. Particularly from (\ref{R3}) we have that
\begin{equation}  \label{n4}
d \tilde{\overline{\lambda}}_{AB} = d \tilde{\lambda}_{AB} = q_{AB}(f)\, df
\end{equation}
where $q_{AB}(f)$ is derived from $d\tilde{\overline\lambda}_{AB}$ as indicated in proposition 3.

\paragraph{Case (a):} From proposition \ref{pr2} in Appendix B and equations (\ref{e17c}) and (\ref{n4}), we have that
\begin{equation}  \label{n5}
\overline{R}_3 = 0
\end{equation}
which is a necessary condition to be fulfilled by $g_{ab}$ in order to admit a degenerate deformation law.
Thus we must first check whether $\overline{R}_3= 0$ and then take $q=0$ if $\det(q_{AB})=0$ or $q={\rm sign}\,(\det(q_{AB}))$ otherwise.
\begin{list}
{(a.\arabic{llista})}{\usecounter{llista}}
\item If $q=0$ we have that [equation (\ref{R141aa})] $\tilde{\overline\lambda}_{AB}=\tilde{\lambda}_{AB}= \hat{\lambda}_{AB} + \hat q_{AB} F$, with $\hat\lambda_{AB}$ and $\hat q_{AB}$ constant, $\hat\lambda^{AB} \hat q_{AB}=0$ and $\det(\hat\lambda_{AB}=-1$. Hence, from $d\tilde{\overline\lambda}_{AB} = \hat q_{AB}\, dF$ it is immediate to determine $\hat q_{AB}$ and $dF$ (appart form a constant factor).

Now, by (\ref{R141aa}) we also have that $\tau=\tau_0$ constant and $D_b F_c=0$, which leads to 
\begin{equation} \label{n6}
a= \tau_0/\overline\tau \qquad {\rm and} \qquad \overline D_b F_c - \psi_{(b} F_{c)} + \frac12\,(\psi_e F_a \overline h^{ae}) \overline h_{bc} = 0
\end{equation}
where $\psi:=\log \varphi$ and the relation between both connections, $D$ and $\overline D$ has been included.

On the one hand, the second equation implies that 
\begin{equation} \label{n7}
 \overline h^{bc} \overline D_b F_c  = 0
\end{equation}
which is a constraint on $F$ and, on the other, it allows to obtain
\begin{equation} \label{n8}
 \psi_b = \overline D_b \log \| F \|^2 \,, \qquad {\rm where} \qquad \log \| F \|^2 := F_e F_a \overline h^{ae} \,, 
\end{equation}
that is, $\psi = \log \| F \|^2 +\,$constant.

Combining now this equation with (\ref{e17a}) and including that $h_{ab} = e^\psi \overline h_{ab}$,
we arrive at \cite{Eisenhart}
\begin{equation} \label{n9}
\overline\mathcal{R} + \overline D^b \overline D_{b}\log \| F \|^2  = 0
\end{equation}
which is a further necessary condition connecting $F$ and $\overline\mathcal{R}$.

\item If $q=-1$ [see Appendix B, right after (\ref{R41})], then $\lambda_{AB}$ must be constant and this implies that  $a\in \Lambda^0\mathcal{S}$ must exist such that $a \overline\lambda_{AB} = \lambda_{AB}=\,$constant. In this case equation (\ref{e17a}) becomes a condition on the conformal factor $\varphi = e^\psi$, namely \cite{Eisenhart} 
$$\overline\mathcal{R} + \overline h^{bc} \overline D_{bc}\psi  = 0$$

\item If $q=+1$, define 
\begin{equation} \label{n10}
df := \sqrt{\det (d \tilde{\overline\lambda}_{AB})} \qquad {\rm and} \qquad q_{ab} := \frac{d \tilde{\overline\lambda}_{AB}}{df}
\end{equation}
Then, using (\ref{R132}), (\ref{R142aa}) and the fact that $\tilde{\overline\lambda}_{AB} = \tilde{\lambda}_{AB}$, we obtain 
\begin{equation} \label{n11}
\tilde{\overline\lambda}_{AB} + q_{AB} = e^f \hat n_A \hat n_B \qquad {\rm and} \qquad 
- \tilde{\overline\lambda}_{AB} + q_{AB} = e^{-f} \hat m_A \hat m_B
\end{equation} 
where $\hat m_A$, $\hat n_A$ are constant and $\hat m_2 \hat n_1 - \hat m_1\hat n_2 = 1$. This is a necessary condition to be fulfilled by $\overline\lambda_{AB}$ which will ensure that (\ref{e17f}) is satisfied and will allow to derive $f$, $\hat m_A$ and $\hat n_A$.

The functions $f$ and $t= \log\tau$ must fulfill (\ref{R15}) and $R_2=0$, which respectively amount to:
\begin{eqnarray} 
\label{n12a}
 & & \overline D^b f_b + \overline h^{bc} t_b f_c  =0 \,, \\
\label{n12b}
 & &  \overline{D}_b \left[ t + \log\|f\|^2 - \psi \right] - \frac1{\|f\|^2} \, (\overline D^a f_a) \, f_b = 0 \\
\label{n12c}
 & & \overline D^b t_b + \|t\|^2  =0 \,, \\
\label{n12d}
 & &  \overline{D}_b \left[ t + \log\|t\|^2 - \psi \right] - \frac1{\|t\|^2} \, ( \overline h^{ac} t_a f_c) \, f_b = 0  \qquad {\rm and}\\
\label{n12e}
 & & \|t\|^2 = \|f\|^2 \qquad {\rm where} \qquad \|t\|^2 :=  \overline h^{ac} t_a t_c
\end{eqnarray} 
Furthermore, the condition (\ref{e17a}) implies that (see \cite{Eisenhart}) $\quad \overline\mathcal{R} + \overline h^{bc} \overline D_c\psi_b  = 0$, or
\begin{equation} \label{n12f}
\overline\mathcal{R} + \overline D^b t_b + \overline D^b \overline D_{c} \log \|t\|^2 + \overline D^b \left(\frac{\overline h^{ae}t_a f_e}{\|t\|^2} \, f_b \right) = 0 
\end{equation}

Since $f$ is known, equations (\ref{n12a}) and (\ref{n12e}) allow to determine 
\begin{equation} \label{n13}
t_b = - \frac{(\overline D^a f_a)}{\|f\|^2} \, f_b \pm \frac{\sqrt{\|f\|^4 - (\overline D^a f_a)^2}}{\|f\|^2} \, \epsilon_b^{\;c} f_c 
\end{equation}

On its turn, equation (\ref{n12d}) is a combination of (\ref{n12a}), (\ref{n12b}) and (\ref{n12e}); equation (\ref{n12b}) yields $\psi_b$ and the remaining two equations, i.e. (\ref{n12c}) and (\ref{n12f}), imply conditions to be fulfilled by $f$, respectively
\begin{equation} \label{n14a}
- \overline D^b \left( \frac{(\overline D^a f_a)}{\|f\|^2} \, f_b \right) \pm \overline D^b \left(\frac{\sqrt{\|f\|^4 - (\overline D^a f_a)^2}}{\|f\|^2}\right) \, \epsilon_b^{\;c} f_c + \|f\|^2 = 0
\end{equation}
and 
\begin{equation} \label{n14b}
\overline\mathcal{R} - \|f\|^2 + \overline D^b \overline D_{c} \log \|f\|^2 - \overline D^b \left( \frac{(\overline D^a f_a)}{\|f\|^2} \, f_b \right)   = 0 
\end{equation}
\end{list}

\paragraph{Case (b):} If the twists $\overline\theta_A$ do not vanish we are compelled to try with case (b) and (\ref{n3}) imposes a first retriction, namely, 
\begin{equation}   \label{n16}
\mbox{a couple of constants $(k_1,k_2)\neq (0,0)$ must exist such that}\qquad \overline\theta_A = k_A \overline\theta
\end{equation}
Furthermore, from proposition 2 in Appendix B we have that $\lambda \in\Lambda^0\mathcal{S}$ and $\hat\lambda_{AB}$ constant must exist such that $\lambda_{AB} = k_A k_B \lambda + \hat\lambda_{AB}$. Taking in consideration (\ref{n2}), this is equivalent to 
\begin{equation}   \label{n17}
\exists a\in \Lambda^0\mathcal{S} \quad \mbox{such that} \quad 
a \overline\lambda_{AB}l^B = p_A \, \mbox{ constant}\,, \quad {\rm with} \quad l^A=(k_2,-k_1)
\end{equation} 
[as $\lambda_{AB}$ is non-degenerated, $(p_1,p_2)\neq (0,0)$], which in turn is equivalent to
\begin{equation}   \label{n18}
 \exists q^A:=(p_2,-p_1) \quad \mbox{such that} \quad \overline\lambda_{AB} q^A l^B = 0
\end{equation} 
If this happens, the factor $a \in \Lambda^0\mathcal{S}$ is
$$ a =\frac{p_1^2 + p_2^2}{\overline\lambda_{AB} p^A l^B} \qquad {\rm with} \qquad 
p^A:=(p_1,p_2) $$

Now, from (\ref{b3}) and (\ref{n3}) we have that
\begin{equation}   \label{n19}
 \varphi = \frac{a^2 \overline\theta \overline\tau}{C}\,, \qquad {\rm with} \qquad C= {\rm constant}
\end{equation} 
This factor must furthermore fulfill the additional conditions implied by (\ref{b4}) and (\ref{b5}) that, after some algebra, yield \cite{Eisenhart}
\begin{eqnarray}
 & & \overline D^b \psi_b + \overline\mathcal{R} + \frac{3 \alpha C \overline\theta a}{\tau^3}\,, \qquad \qquad \overline D^b\tau_b = \frac{\alpha C \overline\theta a}{\tau^2}\,  \label{n20a} \\
& &  {\rm and}  \qquad \psi_b = \overline D_b \log\|\tau\|^2 + \tau_b\,\frac{\alpha C^2 \varphi}{\tau^3 \|\tau\|^2} \label{n20b}
 \end{eqnarray}
where $\psi := \log\varphi$, $\tau = a \overline \tau$ and $\|\tau\|^2 := \tau_b \tau_c \overline h^{bc}$. 

Summarizing, if the twists $\overline\theta_A$ do not vanish we must first check whether (\ref{n16}) and (\ref{n18}) are fulfilled; then compute $a$ and $\varphi$ defined by (\ref{n19}) and check if the relations (\ref{n20a}) and  (\ref{n20b}) hold.

\subsection{Example: stationary axisymmetric spacetimes \label{SS5.1}}
We now consider the case of a stationary axisymmetric spacetime \cite{KSMH2} whose line element is
\begin{equation} \label{ea6a}
ds^2 = e^{-2U+2K}\,\left(d\rho^2 + dz^2\right) + e^{-2U} \rho^2 d\phi^2 - e^{2U}\left( dt + N d\phi\right)^2 
\end{equation}
where $N$, $U$ and $K$ are arbitrary functions of $\rho$ and $z$. The Killing vectors are $X_1=\partial_\phi$ and $X_2 = \partial_t$ and the associated 1-forms are
$$\overline\xi_1 = N \overline\xi_2 + e^{-2U}\rho^2\,d\phi \qquad {\rm and} \qquad \overline\xi_2 = - e^{2U}\,\left(dt + N \,d\phi \right)\,, $$
Therefore we have that $\overline{h}_{bc} = e^{-2U+2K}\,\delta _{bc}$ and 
\begin{equation} \label{ea2}
\overline\lambda_{AB} = e^{-2U} \rho^2 \delta^1_A \delta^1_B - e^{2U} u_A u_B \qquad {\rm with} \qquad u_1 = N\,, \quad u_2=1 \,,
\end{equation}
the determinant is $\overline\tau = \sqrt{2} \,\rho$ and the inverse matrix is
\begin{equation} \label{ea3}
\overline\lambda^{AB} = - e^{-2U}  \delta^A_2 \delta^B_2 + \frac{ e^{2U}}{\rho^2} v^A v^B \qquad {\rm with}\qquad  v^1=1 \,, \quad v^2=- N \,.
\end{equation}

It can be easily checked that
$$ \overline{R}_3 = 0 \qquad \Leftrightarrow \qquad  \overline\lambda^{AB}\,d\overline\lambda_{1A} \wedge d\overline\lambda_{2B} = 0 \qquad \Leftrightarrow \qquad  e^{-2U}\rho = L(N) \,,$$
where $L(N)$ is an arbitrary function of the variable $N$. 

Then, by differentiating $\tilde{\overline\lambda}_{AB} := \overline\lambda_{AB} \sqrt{2}\,/\overline\tau$ we obtain
\begin{equation} \label{ea6}
d \tilde{\overline\lambda}_{AB} = Q_{AB} dN  \qquad {\rm with} \qquad Q_{AB} = L^\prime \delta_A^1 \delta_B^1 + \frac{L^\prime}{L^2} \,u_A u_B - \frac{2}{L}\, \delta_{(A}^1 u_{B)}
\end{equation}
Now, let $Q:=\det(Q_{AB})= (L^{\prime 2} -1)/L^2$. If $Q\neq 0$, we must take $df := dN\,\sqrt{|Q|}$ and $q_{AB}:=Q_{AB}/\sqrt{|Q|}$ and, by (\ref{R15b}) we have that
$$ \frac{dQ_{AB}}{dN} - \frac1{2 Q}\, \frac{dQ}{dN}\, Q_{AB} - |Q|\, \tilde{\overline\lambda}_{AB} = 0 $$
This has a solution only in case that $L^\prime = 0$, which contradicts the initial assumption that $Q\neq 0$. 

If $Q = 0$, by conveniently choosing the sign of $\phi$ we get $L^\prime = 1$ or $L = N + C$ with $C=\,$constant. Then,
the case $q=0$ in Appendix B applies and, from equation (\ref{R141a}), we have that
\begin{equation} \label{ea8a}
\tilde{\overline\lambda}_{AB} = \hat \lambda_{AB} + F \hat q_{AB} \qquad {\rm with} \qquad 
\hat\lambda_{AB} = \left(\begin{array}{cc}
                       2 C & -1 \\  -1 & 0
                         \end{array}  \right)  \,, \qquad 
\hat q_{AB} = \left(\begin{array}{cc}
                       C^2 & -C \\  -C & 1
                         \end{array}  \right)  
\end{equation}
and $F= - L^{-1}$. The results for the case (a.1) in section 4 also apply and we have that
\begin{equation} \label{ea8}
a=\tau_0/\overline\tau \,, \qquad \qquad 
\psi = 6 U - 2 K + \log H \,, \qquad  \qquad  \delta^{bc} \partial_{bc} \left(e^{2U} /\rho \right) = 0
\end{equation}
with $H := \rho^{-4} \delta^{bc} \left(2 \rho U_b -\delta^1_b \right)\left(2\rho U_c -\delta^1_c \right)$.

Besides, we must take in consideration that $h_{bc} := e^\psi \overline h_{bc} = e^{\psi - 2U + 2 K} \delta_{bc}$ is flat, which is equivalent to \cite{Eisenhart}
\begin{equation} \label{ea9}
 \delta^{bc} \partial_{bc}\left(4 U + \log H \right) = 0
\end{equation}

Summarizing, a degenerated deformation law exists that transforms the stationary axisymmetric metric (\ref{ea6a}) into a flat metric iff:
(i) $\overline R_3 =0$, (ii) a constant $C$ exists such that $\rho e^{-2U} = N + C$, (iii) $\overline\lambda_{AB}/\rho$ has the form (\ref{ea8a}) 
and (iv) $U$ simultaneously fulfills (\ref{ea8}) and (\ref{ea9}).
In such a case, the biconformal factors are $a = \tau_0/\overline\tau$ and $\varphi=e^\psi$ with $\psi$ given by (\ref{ea8}).


\section*{Acknowledgements}
 J.C.  acknowledges financial support from the
Spanish Ministerio de Educación y Ciencia through grant No. FPA-2007-60220.
Partial financial support from the Govern de les Illes Balears is
also acknowledged. J.Ll. acknowledges financial support from
Ministerio de Educación y Ciencia through grant No. FIS2007-63034 and from the
Generalitat de Catalunya, 2009SGR-417 (DURSI).

\subsection*{Appendix A: Some bivectors and derivatives}
The following bivectors and bivector equalities will be useful:
\begin{equation} \label{a1}
\Omega_{ab}:=\frac{\tau}{2}\,\xi^1_a \wedge\xi^2_b = \frac{\tau}{2}\,\sigma_{AB}\xi^A_a \xi^B_b =\frac1\tau\,\sigma^{CD}\xi_{Ca} \xi_{Db} = - \frac1\tau\,\xi_{1a} \wedge \xi_{2b} \,,
\end{equation}
where $\sigma_{AB}$ is skewsymmetric and $\sigma_{12}=1$. It is obvious that
$$ \Omega_{ac}\Omega^{ab}= -\frac12\,\xi^B_c \xi_B^b
\qquad {\rm and} \qquad \sigma^{AB} = \lambda^{AC}\lambda^{BD}\sigma_{CD}\,\frac{\tau^2}{2}\,,
\qquad \sigma^{AB} \sigma_{BC} = \delta^A_C$$
where  $\sigma^{12}=-1$, and that $\mathcal{L}_{X_A}\Omega_{ab} =0$.

The volume tensor on ${\cal S}$:
\begin{equation} \label{a2}
\epsilon_{ab}:=\frac{\sqrt{2}}{\tau}\,\epsilon_{abcd}X_1^c X_2^d = -\frac1{\tau\sqrt{2}}\,\sigma^{CD} \epsilon_{abcd}X_C^c X_D^d
\end{equation}
and $\epsilon^{ab} \epsilon_{cb} = h^a_c $. It is obvious that $\epsilon_{cb}$ is horizontal and Lie-constant, hence $\epsilon_{cb}\in\Lambda^2{\cal S}$.

The dual bivectors respectively are:
\begin{equation} \label{a3}
\tilde{\epsilon}_{ef}= \sqrt{2}\Omega_{ef} \,, \qquad \tilde{\Omega}_{ef}= -\frac1{\sqrt{2}}\,\epsilon_{ef}
\end{equation}
Furthermore, if $w^a$ is a vector on ${\cal S}$, then
\begin{equation}  \label{a4}
\left(\xi^A \wedge w \right)^\sim_{ef} = - \frac{2\sqrt{2}}{\tau}\,\sigma^{AB}\,\xi_{B[e} \epsilon_{f]c} w^c
\end{equation}

If $w^b$ is a vector field on ${\cal S}$,  from $[X_A,w]=0$ it follows that
\begin{equation}  \label{a5}
\nabla_A w^b = w^d\,\left(\frac12\,\lambda_{AB|d} \xi^{B\,b} + \frac12\,\theta_A\,\epsilon_d^{\;\;b}  \right)
\end{equation}
and also,
\begin{equation}  \label{a5a}
\nabla_A T^{bc} = T^{dc}\,\left(\frac12\,\lambda_{AB|d} \xi^{B\,b} + \frac12\,\theta_A\,\epsilon_d^{\;\;b}  \right) + T^{bd}\,\left(\frac12\,\lambda_{AB|d} \xi^{B\,c} + \frac12\,\theta_A\,\epsilon_d^{\;\;c}  \right)
\end{equation}
where  $\nabla_A$ stands for $X_A^a\nabla_a$.  Now, using the identity: $d(\log|det\lambda_{AB}|)= d\lambda_{AB} \lambda^{AB}$, 
from (\ref{e2}) we have
\begin{equation} \label{a6}
t_a = \frac12\, \lambda^{AB}\,\lambda_{AB|a} \,, \qquad t:=\log\,\tau    
\end{equation}

\section*{Appendix B}
Our goal here is to see how equations (\ref{e20d}),  namely
$$ R_1 = R_2 = R_3 \,,\qquad P_{Ab} = Q_{Ab}=0 \,, \qquad P_{DcAb} = 0 $$
constrain the possible values of $\lambda_{AB}$, $\kappa^A_a$ and $h_{ab}$,
\begin{proposition}
If $P_{Ab} = Q_{Ab}=0$ , then
\begin{list}
{(\alph{llista})}{\usecounter{llista}}
\item  either $\theta_1 = \theta_2 =0$ or
\item two constants, $k_A\,$, exist such that: $\theta_A = k_A \theta$ and $d \lambda_{AB} = k_A k_B d\lambda$, where $\theta,\,\lambda \in\Lambda^0(\mathcal{S})$. In this  case one also has 
\begin{equation}  \label{b3}
\theta \tau = C \qquad {\rm and} \qquad -\frac{\tau^2}{2} = \alpha \lambda + \hat \delta
\end{equation}
with $C$, $\alpha$ and $\hat\delta$ constant.
\end{list}
\end{proposition}

\paragraph{Proof:} Indeed, by (\ref{e17d}) $P_{Ab}=0$ implies $2 d\theta_A + \theta_C \lambda^{CT} d\lambda_{TA} =0$. Multiplying it by $\theta_{A^\prime}$, $A^\prime\neq A$, and using that $Q_{Tb}=0$ amounts to $\theta_1 d\lambda_{T2} = \theta_2 d\lambda_{T1}$, one readily obtains that $\theta_1 d\theta_2 = \theta_2 d\theta_1$, which implies 
\begin{list}
{(\alph{llista})} {\usecounter{llista}}
\item either $\theta_1 = \theta_2 =0$ or
\item two constants $k_A$ exist such that $\theta_A = k_A \theta$, with $\theta\in\Lambda^0(\mathcal{S})$.

Furthermore, substituting this into $\theta_1 d\lambda_{A2} = \theta_2 d\lambda_{A1}$ and taking the symmetry of $\lambda_{AB}$ into account, we obtain that $ d\lambda_{AB} = k_A k_B d\lambda$, with $\lambda\in\Lambda^0\mathcal{S}$, and therefore
\begin{equation} \label{b1}
\lambda_{AB} = k_A k_B \lambda + \hat{\lambda}_{AB} \qquad {\rm with} \qquad \hat{\lambda}_{AB} =  \hat{\lambda}_{BA} = {\rm constant}
\end{equation}
The inverse matrix $\lambda^{AB}$ is
\begin{equation} \label{b2}
\lambda^{AB} = \frac{-2}{\tau^2}\,\left(l^A l^B \lambda + \hat{\lambda}^{AB}\right) \quad {\rm where} \quad l^A= (k_2,\, -k_1) \,,  \quad
\hat{\lambda}^{AB} = \left(\begin{array}{cc}
               \hat{\lambda}_{22} & -\hat{\lambda}_{12}\\  -\hat{\lambda}_{12} & \hat{\lambda}_{11}
               \end{array} \right)
\end{equation}
and we also have that
\begin{equation} \label{b2a}
- \frac{\tau^2}{2}=\det(\lambda_{AB})= \alpha \lambda + \hat\delta\,,\qquad {\rm with} \qquad \alpha=\hat{\lambda}^{AB} k_A k_B\qquad {\rm and} \qquad \hat\delta := \det(\hat{\lambda}_{AB})\,.
\end{equation}
Substituting these into $P_{Ab}=0$, we obtain $2 d\theta + \theta k_C \lambda^{CT} k_T d\lambda= 0$, which implies that
$d\left(\theta \left|\det(\lambda_{AB})\right|^{1/2}\right) = 0 \quad {\rm or} \quad \theta \tau = C \,$,  constant,
where (\ref{e2}) has been included. \hfill $\Box$
\end{list}

Let us now study the implications of the remaining curvature equations, $R_1=R_2=R_3=0$ and $P_{DcAb}=0$.
Consider first the case (b): $\theta_A = k_A \theta$ and $d \lambda_{AB} = k_A k_B d\lambda$. Equations $R_2=0$ and $R_3=0$ 
are identically satisfied and do not imply any further condition on $\theta$, $\tau$ or $k_A$. Then, taking (\ref{b2}) and (\ref{b3}) into account, equation $R_1=0$ implies that
\begin{equation}  \label{b4}
\mathcal{R} = -\,\frac{3 \alpha C^2}{ \tau^4}
\end{equation}
and equation $P_{DcAb} =0$ amounts to
\begin{equation}  \label{b5}
D_{bc} \tau + \frac{\alpha C^2}{2 \tau^3}\, h_{bc} =0 
\end{equation}
This is a partial differential system that is integrable provided that the Ricci scalar is (\ref{b4}). Combining now equations (\ref{b4}) and (\ref{b5}) we arrive at
\begin{equation}  \label{b5a}
D_{bc} \mathcal{R}^{-1/4}  +  \frac{1}{6}\,\mathcal{R}^{3/4}\, h_{bc} = 0
\end{equation}
which is a condition to be fulfilled by $\mathcal{R}$, the Ricci scalar of the given metric $h_{ab}$ on $\mathcal{S}$, in order that the ambient flat metric $\eta_{ab}$ exists.

Consider now the case (a): $\theta_B=0$, $B=1,2$, i.e. orthogonal transitivity, with no restriction on $\lambda_{AB}$. To start with, by  (\ref{e17a}), equation $R_1=0$ amounts to $\mathcal{R}=0$, which implies that $h_{ab}$ is flat.

\begin{proposition}  \label{pr2}
The scalar $R_3$ vanishes if, and only if, a function  $f\in\Lambda^0\mathcal{S}$ and functions $q_{AB}(f)$ exist such that
\begin{equation} \label{R3}
d\left(\sqrt{2}\, \tau^{-1} \lambda_{AB}\right) =  q_{AB}(f)\, df  \,, \qquad \lambda^{AB}q_{AB} =0 \qquad{\rm and} 
\quad q:=\det(q_{AB}) \in \{0,\,\pm 1\}
\end{equation}
\end{proposition}

\paragraph{Proof:} Define $\tilde{\lambda}_{AB} :=\sqrt{2}\, \tau^{-1} \lambda_{AB} $. It is obvious from (\ref{e2}) that $\det(\tilde{\lambda}_{AB}) = -1$. From (\ref{e17c}) we have that
$R_3 = 2^{-3/2}\,\epsilon^{bc}\tilde\lambda^{TE} \tilde\lambda_{1E|c} \tilde\lambda_{2T|b} $, where $\tilde\lambda^{AB} \tilde\lambda_{BC}= \delta^A_C$, and therefore,
$$R_3=0 \qquad \Leftrightarrow \qquad  \tilde\lambda^{ET} \,d\tilde{\lambda}_{1E} \wedge d\tilde{\lambda}_{2T} = 0 $$
A short calculation then proves that this is equivalent to the existence of $F\in\Lambda^1\mathcal{S}$ such that $d\tilde{\lambda}_{AB} \propto F$. Now, since dim$\mathcal{S}=2$, $F$ is integrable, i.e. proportional to $du$ for some $u\in \Lambda^0\mathcal{S}$, whence it follows that $d\tilde\lambda_{AB} = Q_{AB} du$ for some $Q_{AB}\in\Lambda^0\mathcal{S}$ and the integrability conditions imply that $Q_{AB}= Q_{AB}(u)$.

Then (\ref{R3}) follows taking $f=u$ and $q_{AB} = Q_{AB}$, if $Q(u):=\det(Q_{AB}) = 0$, or taking 
$df=\sqrt{|Q|} \,du$ and $q_{AB} = Q_{AB}/\sqrt{|Q|}$, if $Q(u)\neq 0$.   
Furthermore,  $\det(\tilde{\lambda}_{AB}) = -1$ implies that $\tilde\lambda^{AB} d\tilde\lambda_{AB}= 0$ or $\tilde\lambda^{AB} q_{AB}= 0$.
\hfill $\Box$

Notice that neither $f$ nor $q_{AB}$ vanish except in the trivial case $\tilde\lambda_{AB} =$constant. 

$q_{AB}$ is a symmetric, traceless, 2-square matrix of functions on $\mathcal{S}$. Since the number of dimensions is 2, using the characteristic polynomial we have that
\begin{equation} \label{R4}
q_{AB}\tilde\lambda^{BC}q_{CD}= q \tilde\lambda_{AD} \,, \qquad q:= \det(q_{AB}) \in\{0,\,\pm1\} \,.
\end{equation}

Consider now the following quadratic form on the space of symmetric 2-square matrices:
\begin{equation} \label{R6}
m_{AB}  \,\longrightarrow m_{AB} m_{CD}\tilde\lambda^{AC} \tilde\lambda^{BD} = \left(m_{CD}\tilde\lambda^{CD}\right)^2 +  \det(m_{CD})
\end{equation}
It can be easily seen that it is non-degenerate and has signature$(+\,+\,-)$. 
We can then complete a base $\{\tilde\lambda_{AB},\,q_{AB}, \, w_{AB} \}$ in this space of symmetric matrices such that
\begin{equation} \label{R5}
\left.  \begin{array}{lll}
q_{AB}\tilde\lambda^{AB} = 0 \,,&  q_{AB}\tilde\lambda^{AC}\tilde\lambda^{BD} q_{CD}= 2 q  \,,& \\[1ex]
w_{AB}\tilde\lambda^{AB} = 0 \,,&  w_{AB}\tilde\lambda^{AC}\tilde\lambda^{BD}w_{CD}= -2 q  \,,& 
 w_{AB}\tilde\lambda^{AC}\tilde\lambda^{BD}q_{CD}= 2(|q|-1) 
\end{array}  \right\}
\end{equation}
and, besides, $\det(q_{AB}) = -\det(w_{AB}) = q$ and $ \det(\tilde\lambda_{AB})= -1$. 

$\{\tilde\lambda_{AB},\,q_{AB}, \, w_{AB} \}$ is thus a rigid base for the quadratic form (\ref{R6}): an orthogonal base in the case $q\neq 0$  and a base containing two conjugate null vectors in the case $q=0$. In all instances, $w_{AB}$ is thoroughly determined by $\tilde\lambda_{AB}$ and $q_{AB}$. We thus have the following differential equations [the first one comes from (\ref{R3})]:
\begin{equation} \label{R10}
\left. \begin{array}{ll}
       \displaystyle{\frac{d\tilde\lambda_{AB}}{df} =  q_{AB} } \,, \qquad &
       \displaystyle{\frac{dq_{AB}}{df} = q \tilde\lambda_{AB} + q_{AB} \,(1-|q|) u  + q w_{AB} \,v } \\
       & \displaystyle{\frac{dw_{AB}}{df} = (|q|-1) \tilde\lambda_{AB} + q \, q_{AB}\, v - w_{AB} \,(1-|q|) u}
      \end{array}
       \right\}
\end{equation}
where $q= \pm 1$ or $0$ and $u(f)$ and $v(f)$.

As the quadratic form (\ref{R6}) can be associated with a non-degenerate metric product in the 3-space of symmetric 2-square matrices, these equations can be seen as a sort of ``Frênet-Serret equations''.

In the case $q=0$, they yield
\begin{equation} \label{R10a}
\frac{d\tilde\lambda_{AB}}{df} =  q_{AB}  \,, \qquad \frac{dq_{AB}}{df} = u\, q_{AB} \,, \qquad 
\frac{dw_{AB}}{df} = - \tilde\lambda_{AB} - u\, w_{AB} 
\end{equation}
whose solution is 
\begin{equation}  \label{R131}
 \tilde\lambda_{AB} = \hat\lambda_{AB} + F \hat{q}_{AB} \,, \qquad q_{AB} = \dot F \hat{q}_{AB}  
\end{equation}
where $\hat{\lambda}_{AB}$ and $\hat{q}_{AB}$ are constant matrices satifying (\ref{R5}), $F=F(f)$ and $\dot F:=dF/df$;
whereas in the case $q=\pm 1$  (\ref{R10}) reads
\begin{equation} \label{R10b}
\frac{d\tilde\lambda_{AB}}{df} =  q_{AB}  \,, \qquad \frac{dq_{AB}}{df} = q \tilde\lambda_{AB} + q v\, w_{AB} \,, \qquad 
\frac{dw_{AB}}{df} = q v\, q_{AB} 
\end{equation}

\paragraph{\underline{The condition $R_2=0$}} 
From (\ref{R3}) and (\ref{e17b}) it easily follows that
\begin{equation} \label{R41}
R_2 = \frac12\,h^{bc}\,\left(t_b t_c - q f_b f_c\right) 
\qquad {\rm with} \qquad t:=\log \tau \,,
\end{equation}
where (\ref{R4}) and (\ref{R5}) have been used.
Now, as $h_{bc}$ is positive definite, the condition $R_2=0$ allows two different cases:
\begin{equation} \label{R41a}
\left.  \begin{array}{ll}
\mbox{if }q=0\,,\mbox{ then } & t_b =0  \mbox{ and } \tau = {\rm constant} \\
\mbox{if }q=1\,,\mbox{ then } & h^{bc} t_b t_c = h^{bc} f_b f_c  
\end{array}   \right\}
\end{equation}
The case $q=-1$ is forbidden, because $R_2=0$ would imply $f_b=0$ and $d\tilde\lambda_{AB}= 0$, which amounts to $q_{AB}=0$, in contradiction with $det(q_{AB})=-1$.

\paragraph{\underline{The condition $P_{DcAb}=0$}}
\begin{description}
\item [For $q=0$,] on account of (\ref{R131}) and (\ref{R41a}), $P_{DcAb}=0$ implies that $ D_b F_c = 0$. Hence, 
using (\ref{R131}) and (\ref{R41a}), we have that  
\begin{equation} \label{R141a}
\lambda_{AB} = \frac{\tau_0}{\sqrt{2}}\,\left[  \hat\lambda_{AB} +  F\,\hat{q}_{AB} \right]
\qquad {\rm with}  \qquad D_{ab}F =0 \qquad {\rm and} \qquad \tau = \tau_0 ={\rm constant}
\end{equation}

\item[For $q=1$,] substituting (\ref{R10b}) into (\ref{e17f}) and using (\ref{R3}), (\ref{R4}) and (\ref{R5}), we obtain that $P_{DcAb}=0$ amounts to:
\begin{equation} \label{R15}
\mathcal{T}_{bc}:= D_b t_c + \frac{1}2\,t_b t_c + \frac{1}2\, f_b f_c = 0 \,, \qquad
\mathcal{F}_{bc}:= D_b f_c + t_{(b} f_{c)}  = 0 
\end{equation}
supplemented with $v = 0$. 

Using this, equations (\ref{R10b}) read:
\begin{equation} \label{R15b}
\frac{d\tilde\lambda_{AB}}{df} =  q_{AB}  \,, \qquad \frac{dq_{AB}}{df} = \tilde\lambda_{AB} \,, \qquad 
\frac{dw_{AB}}{df} = 0  
\end{equation}
whose solution is
\begin{equation} \label{R132}
\tilde\lambda_{AB} = \hat\lambda_{AB} \cosh f + \hat{q}_{AB} \sinh f
\end{equation}
where $\hat{\lambda}_{AB}$ and $\hat{q}_{AB}$ are constant matrices satifying (\ref{R5}). 

Now, equation (\ref{R15}) is a partial differential system where all the derivatives of the unknowns $t$ and $f$ are specified. The subsequent integrability conditions do not imply any new condition. Moreover, equation (\ref{R41}) implies a further restriction
\begin{equation}  \label{R14}
\frac12\,h^{bc}\left(t_b t_c -  f_b f_c \right) = 0
\end{equation}
which is compatible with (\ref{R15}); indeed, $D_a R_2 + t_a R_2 = 0$, and provided that $R_2$ vanishes at $x_0\in\mathcal{S}$ it vanishes in some open neighbourhood of $x_0$.

We now introduce the new variables $\Phi_\pm := e^{(t \pm f)/2}$ and equations (\ref{R15}) and (\ref{R14}) become
\begin{equation}  \label{R144a}
 D_{bc} \Phi_\pm  = 0 \qquad {\rm and} \qquad h^{bc} D_{b}\Phi_+ D_c \Phi_- =0
\end{equation}
From (\ref{R132}) it then follows immediately that 
\begin{equation} \label{R142}
\lambda_{AB} = 2^{-3/2}\,\left( \Phi_+^2 [ \hat\lambda_{AB} + \hat{q}_{AB}] + \Phi_-^2 [ \hat\lambda_{AB} - \hat{q}_{AB}]\right)  \qquad {\rm and} \qquad
\tau = \Phi_+ \Phi_- 
\end{equation} 
With a little of algebra it can be seen that, as a consequence of (\ref{R5}), there exist $\hat m_A$ and $\hat n_A$ such that 
$\hat m_2 \hat n_1 -\hat m_1 \hat n_2 = 1$ and that
\begin{equation} \label{R142aaa}
\hat\lambda_{AB} + \hat q_{AB}= 2 \hat n_A \hat n_B  \qquad  {\rm and} \qquad \hat\lambda_{AB} - \hat q_{AB}= -2 \hat m_A \hat m_B
\end{equation} 
Therefore, (\ref{R142}) finally yields
\begin{equation} \label{R142a}
\lambda_{AB} = \frac1{\sqrt{2}}\,\left( - \Phi_-^2 \hat m_A \hat m_B + \Phi_+^2 \hat n_A \hat n_B \right) 
\end{equation}
with $\Phi_\pm $ fulfilling (\ref{R144a}) 
\end{description}

\subsection*{Summary: How to proceed? Guidelines}
We start from a given Riemannian metric $h_{ab}$ on $\mathcal{S}$.
\begin{description}
\item[1.-] If $h_{ab}$ is flat, then we take $\theta_A=0$ and
\begin{list}
{{\bf 1.\alph{llista}.-}}{\usecounter{llista}}
\item choose two matrices $\hat\lambda_{AB}, \,\hat{q}_{AB}$  fulfilling (\ref{R5}), with
$q:= \det(\hat{q}_{AB})$
\item then, $\lambda_{AB}$ is given by (\ref{R142a}) if $q=1$ or by (\ref{R141a}) if $q=0$.
\end{list}
\item[2.-] If $h_{ab}$ is not flat, we first check whether $\mathcal{R}$ fulfills (\ref{b5a}). If so, we choose two constants $\alpha \neq 0$ and $C\neq 0$ and take [equation (\ref{b4})]
$$ \tau = \left(-\,\frac{3\alpha C^2}{\mathcal{R}} \right)^{1/4} \,, \qquad \theta = \frac{C}{\tau} = \left(-\,\frac{ C^2 \mathcal{R}}{3\alpha} \right)^{1/4} $$
then choose $k_A$ and $\hat{\lambda}_{AB}$ such that $\hat{\lambda}^{AB} k_A k_B = \alpha$ [equation (\ref{b2a})] and take
$$ \theta_A = k_A \theta\,, \qquad \theta^A = -\,\frac{2C}{\tau^3}\,\hat{\lambda}^{AB} k_B \,, \qquad \lambda_{AB}= k_A k_B \lambda + \hat{\lambda}_{AB} \qquad 
{\rm with} \qquad \lambda :=-\frac1{\alpha}\,\left( \frac{\tau^2}{2} + \delta_0\right) $$
\end{description}

In both cases the covectors $\xi^A \in \Lambda^1\mathcal{M}$ can be determined as indicated in subsection \ref{SS2.4}, i. e. by solving equations (\ref{e20b}) and (\ref{e20c}).

\section*{Appendix C}
\begin{proposition}  \label{Pz1}
The covectors $m_a$ and $n_a$ in the expression (\ref{E1}) for the 2-dimensional elliptic projector $H_{ab}$ can be chosen so that $\mathcal{L}_{X_A} m_a = \mathcal{L}_{X_B} n_a = 0$
\end{proposition}

Indeed, let $m^\prime_a$ and $\,n^\prime_b$ be a couple of covectors fulfilling (\ref{E1}). As
they are determined except for a rotation $\zeta \in \Lambda^0\mathcal{M}$, the couple of covectors
\begin{equation}  \label{A1}
  m_a= m^\prime_a\,\cos\zeta + n^\prime_a\,\sin\zeta \qquad {\rm and} \qquad  n_a= - m^\prime_a\,\sin\zeta + n^\prime_a\,\cos\zeta
\end{equation}
also fulfill (\ref{E1}).
Since $\mathcal{L}_{X_A} H_{ab} = 0$, there exist $\gamma_A\in \Lambda^0\mathcal{M}$, $A=1,2$, such that $\mathcal{L}_{X_A} m^\prime_a = \gamma_A n^\prime_a$ and $ \mathcal{L}_{X_B} n^\prime_a = - \gamma_B m^\prime_a$.
We then choose for $\zeta$ any solution of $X_A \zeta = - \gamma_A$, $A=1,2$,  (it is integrable because
$[X_1,X_2]=0$ implies that $X_A \gamma_B - X_B \gamma_A = 0$) which, substituted in (\ref{A1}), yields $\mathcal{L}_{X_A} m_a = \mathcal{L}_{X_B} n_a = 0$. \hfill $\Box$

From the definitions of $m_A$ and $n_A$, and the fact that both Killing vectors commute, it follows immediately that $\mathcal{L}_{X_A} m_B = \mathcal{L}_{X_A} n_B = 0$, $A,B=1,2$.

We shall thus take
\begin{equation} \label{E1a}
\mu_a := m_a - m_A \overline\xi^A_a \qquad {\rm and} \qquad \nu_a := n_a - n_A \overline\xi^A_a
\end{equation}
and it is then easy to prove that $\mathcal{L}_{X_B} \mu_a = \mathcal{L}_{X_B} \nu_a = 0$.

There is still left the residual freedom to rotate an angle $\tilde\zeta $ such that $X_A \tilde\zeta = 0$, i. e. $\tilde\zeta \in \Lambda^0\mathcal{S} $, which can be used to show that
\begin{proposition}  \label{Pz2}
The covectors $m_a$ and $n_a$ in  (\ref{E1}) can be chosen so that $\overline\lambda^{AB} m_A  n_B = 0$, where $m_A:=m_a X^a_A$ and $n_B:=n_a X^a_B$.
\end{proposition}

We have thus proved that two $g$-orthonormal covectors $m_a$ and $n_a$ can be found such that
\begin{equation} \label{A2}
H_{ab} = m_a m_b + n_a n_b \,,\qquad {\rm with} \qquad \mathcal{L}_{X_A} m_a = \mathcal{L}_{X_A} n_a = 0 \qquad {\rm and} \qquad \overline\lambda^{AB} m_A  n_B = 0
\end{equation}
As a consequence, the covectors $\mu_a$ and $\nu_a$ in (\ref{E1a}) fulfill
\begin{equation} \label{A3}
\mu_a \mu_b \overline{h}^{ab} = 1 -x \,, \qquad \nu_a \nu_b \overline{h}^{ab} = 1 -y
\qquad {\rm and} \qquad  \mu_a \nu_b \overline{h}^{ab} = 0
\end{equation}
where $x:=\overline\lambda^{AB} m_A  m_B$ and $y:=\overline\lambda^{AB} n_A  n_B$. As $\overline\lambda_{AB}$ is hyperbolic and $\overline{h}_{ab}$ is elliptic,  $m_a$ and $n_a$  can be chosen so that $x<0<y<1$. Furthermore we have that\footnote{Assuming the generic case $x y \neq 0$}
\begin{equation} \label{A3a}
\overline\lambda_{AB} = \frac1{x}\, m_A m_B + \frac1{y}\, n_A n_B
\end{equation}

Then it easily follows that $(m_1n_2-m_2n_1)^2= xy \det(\overline\lambda_{AB})$ and, appropriately choosing the signs of $m_A$ and $n_B$, we have that
\begin{equation} \label{A3A}
m_1n_2-m_2n_1= -  \overline\tau \sqrt{-2xy} \qquad {\rm and} \qquad m^1 n^2 - m^2 n^1 = \frac{\overline\tau}{\tau^2}\, \sqrt{-2xy}
\end{equation}

Substituting now (\ref{E1a}) into (\ref{E1}), we obtain the splitting of $H_{ab}$ in its vertical, horizontal and cross components:
\begin{equation} \label{A4}
H_{ab} = H_{AB} \overline\xi^A_a \overline\xi^B_b + \mu_a \mu_b + \nu_a \nu_b + 2 \overline\xi^A_{(a} \left(m_A \mu_{b)} + n_A \nu_{b)} \right)
\end{equation}
where $H_{AB} := H_{ab} X^a_A X^b_B = m_A m_B + n_A n_B\,$.

From equations (\ref{e23}) and (\ref{A4}) it  readily follows
\begin{equation} \label{A5}
\xi_{Aa} = a \overline\xi_{Aa} + b\,H_{AB}\overline\xi^B_a + b \,\left(m_A \mu_{a} + n_A \nu_{a} \right)
\end{equation}
and, taking (\ref{A3a}) and (\ref{A4}) into account, we have that the vertical metric is
\begin{equation} \label{A6}
\lambda_{AB} = a \overline\lambda_{AB} + b H_{AB} = \left(\frac{a}{x} + b \right)\,m_A m_B + \left(\frac{a}{y} + b \right)\, n_A n_B  \,,
\end{equation}
the inverse of which is 
$\lambda^{AB} = \displaystyle{- \frac2{\tau^2}\,\sigma^{AC}\sigma^{BD}\lambda_{CD}}$ where $\sigma^{AB} = - \sigma^{BA}$ and $\sigma^{12}=-1$. Moreover, since $\lambda_{AB}$ is hyperbolic and $a$, $a+b$ and $y$ are positive, it must be $a/x+b <0$, which implies that $-a/b <x$.

It follows immediately from (\ref{A5}) that the shift covector (\ref{p2}) is
\begin{equation} \label{A7}
\kappa^A_a = b \lambda^{AB} \,\left(m_B \mu_{a} + n_B \nu_{a}  \right)
\end{equation}

Comparing the expression (\ref{A6}) for $\lambda_{AB}$ with equation (\ref{F2})
we have that it exists $\zeta\in\Lambda^0\mathcal{S}$ such that
\begin{equation} \label{A7a}
\left. \begin{array}{l}
       \displaystyle{m_A\,\sqrt{-\left(\frac{a}{x}+b\right)} = 2^{-1/4}\,\left( \Phi_- \cosh\zeta\, \hat{m}_A +\Phi_+ \sinh\zeta \,\hat{n}_A  \right)    } \\
 \displaystyle{n_A\,\sqrt{\frac{a}{y}+b} = 2^{-1/4}\,\left( \Phi_- \sinh\zeta \,\hat{m}_A + \Phi_+ \cosh\zeta \,\hat{n}_A  \right)   }
       \end{array}  \right\}
\end{equation}
which, substituted into  into equation (\ref{A3a}), leads to
\begin{eqnarray*}    \label{F3}    
\overline\lambda_{AB} & = & \hat m_A \hat m_B \,\frac{\Phi_-^2}{\sqrt{2}}\, [A_+ \cosh(2\zeta) + A_-] +  
\hat n_A \hat n_B \,\frac{\Phi_+^2}{\sqrt{2}}\, [A_+ \cosh(2\zeta) - A_-] +\\[1ex]
 & & [\hat m_A \hat n_B + \hat m_B \hat n_A] \frac{\Phi_+ \Phi_-}{\sqrt{2}}\, A_+ \sinh(2\zeta) 
\end{eqnarray*}
with
$$ 2 A_\pm := - \frac1{a+bx} \pm \frac1{a+by} $$

From (\ref{A3a}) and (\ref{A6}) it easily follows that
\begin{equation} \label{F5}
\frac1{(a+bx)(a+by)} = \frac{\overline\tau^2}{\tau^2} \qquad {\rm and} \qquad \overline\lambda_{AB}\lambda^{AB} = \frac1{a+bx} + \frac1{a+by}
\end{equation}
As a consequence, $z_1=(a+bx)^{-1}$ and $z_2=(a+by)^{-1}$ are the solutions to the equation \\
$z^2 - \overline\lambda_{AB}\lambda^{AB} \,z + \overline\tau^2/\tau^2 = 0$. Therefore, assuming $b>0$, we have that $z_1 > z_2$ and
\begin{equation} \label{F6}
z_{1,2} = \frac12\, \left(\overline\lambda_{AB}\lambda^{AB} \pm \sqrt{\left(\overline\lambda_{AB}\lambda^{AB}\right)^2 - 4 \overline\tau^2/\tau^2} \right)
\end{equation}
where $\overline\lambda_{AB}\lambda^{AB}$ has to be understood as [see equations (\ref{R142}) and  (\ref{A6})]
$$\overline\lambda_{AB}\lambda^{AB} = - \sqrt{2}\,\overline\lambda_{AB}\,\sigma^{AC} \sigma^{BD}\,\left( - \Phi_+^{-2} \hat m_C \hat m_D + \Phi_-^{-2} \hat n_C \hat n_D \right)  $$
Hence, equations (\ref{F6}) permit to derive 
\begin{equation}  \label{F6ab}
x = \frac{1}{b}\,\left(\frac{1}{z_1} - a \right)   \qquad {\rm and} \qquad 
y = \frac{1}{b}\,\left(\frac{1}{z_2} - a \right)
\end{equation}
as functions of $a$, $b$ and $\Phi_\pm$. 

An expression of $\zeta$ in terms of these variables is also obtained from (\ref{F3}):  
\begin{equation}  \label{F6ac}
 \sinh(2\zeta) = \frac{\sqrt{2}}{\tau (z_1-z_2)}\, \overline\lambda_{AB}\hat{m}^A \hat{n}^B \,.
\end{equation}

Substituting then (\ref{A4}), (\ref{A6}), (\ref{A7}) and (\ref{e23}) into  (\ref{p1}), we arrive at
\begin{equation} \label{A8}
h_{ab} = a \left[\overline{h}_{ab} + \mu_a \mu_b \,\frac{b}{a+bx} + \nu_a \nu_b\,\frac{b}{a+by}  \right]
\end{equation}
where it has been used that
$$ \lambda^{AB} m_A m_B = \frac{x}{a+bx}\,, \qquad \lambda^{AB} n_A n_B = \frac{y}{a+by} \qquad {\rm and}
\qquad \lambda^{AB} m_A n_B = 0 $$
Now, according to (\ref{A3}), there exists an $\overline{h}$-orthonormal base $\{\hat\mu_a,\, \hat\nu_a\}$ of $T\mathcal{S}$ such that $\mu_a = \hat\mu_a\,\sqrt{1-x} $ and  $\nu_a = \hat\nu_a\,\sqrt{1-y} $ which, substituted into equation (\ref{A8}), yields
\begin{equation} \label{A9}
h_{ab} = a (a+b) \,\left[ z_1 \hat\mu_a \hat\mu_b + z_2 \hat\nu_a \hat\nu_b \right]
\end{equation}
with $z_1$ and $z_2$ given by (\ref{F6}).

\end{document}